\documentclass[a4paper,11pt,fleqn]{article}
\usepackage{amsfonts,epsfig,epic,latexsym}

\newlength{\dinwidth}
\newlength{\dinmargin}
\def\nn{\nonumber}
\def\non{\nonumber\\}
\def\be{\begin{equation}}
\def\ee{\end{equation}}
\def\ben{\begin{displaymath}}
\def\een{\end{displaymath}}
\def\ba{\begin{eqnarray}}
\def\ea{\end{eqnarray}}

\newcommand{\qed}{\begin{flushright}$\Box$\end{flushright}}

\def\D{\Delta}
\def\d{\delta}
\def\e{\varepsilon}
\def\f{\varphi}

\def\g{\gamma}
\def\k{\kappa}

\def\m{\mu}
\def\n{\eta}
\def\O{\Omega}
\def\P{\Psi}

\def\r{\rho}
\def\s{\sigma}
\def\S{\Sigma}
\def\t{\tau}
\def\th{\theta}

\def\x{\xi}
\def\z{\zeta}
\def\e{\epsilon}

\def\A{{\cal{A}}}
\def\cC{{\cal C}}

\def\cL{{\cal L}}
\def\cO{{\cal O}}
\def\U{{\cal U}}

\def\bA{{\mbox{\boldmath{$A$}}}}
\def\cV{{\cal V}}
\def\cAt{\hat{\cal{A}}}

\def\zb{{\bar z}}
\def\xb{{\bar{\xi}}}

\def\gb{{\bar \gamma}}
\def\mb{{\bar \mu}}

\def\C{\mathbb{C}}

\def\R{\mathbb{R}}

\def\la{\label}
\def\ci{\cite}

\def\Ref#1{(\ref{#1})}

\def\f{\frac}

\def\i{\infty}
\def\p{\partial}

\def\tr{{\rm tr}}

\def\coset{SL(2,\R)/SO(2)}

\def\Cx{{\cC}^{\x}}
\def\Cxb{{\cC}^{\xb}}

\def\Psit{\hat{\Psi}}
\def\At{\hat{A}}
\def\Ht{\hat{H}}

\def\ga{\mathfrak{sl}(2,\R)}
\def\G{SL(2,\R)}

\makeatletter
\@addtoreset{equation}{section}
\makeatother

\newtheorem{Theorem}{Theorem}[section]
\newtheorem{Lemma}{Lemma}[section]
\newtheorem{Note}{Note}[section]
\newtheorem{Definition}{Definition}[section]
\newtheorem{Corollary}{Corollary}[section]

\begin{document}
\renewcommand{\thefootnote}{\fnsymbol{footnote}}
\begin{flushright}
DESY 96-130\\
hep-th/9607095\\
July 1996\\
\end{flushright}
\begin{center}
\vspace*{1cm}
{\huge  Quantization of coset space $\s$-models}\smallskip\\
{\huge  coupled to two-dimensional gravity \footnote{to appear in 
{\it Commun. Math. Phys.}}}\bigskip\\ 
{\large D. Korotkin\footnote{On leave of absence from Steklov
Mathematical Institute, Fontanka, 27, St.Petersburg 191011 Russia} and
H. Samtleben\bigskip\medskip\\ { II. Institut f\"ur Theoretische
Physik, Universit\"at Hamburg,}\\ { Luruper Chaussee 149, 22761
Hamburg, Germany}} 
\end{center}
\renewcommand{\thefootnote}{\arabic{footnote}}
\setcounter{footnote}{0}
\vspace*{1cm}
\hrule
\begin{abstract}
The mathematical framework for an exact quantization of the
two-dimensional coset space $\s$-models coupled to dilaton gravity,
that arise from dimensional reduction of gravity and supergravity
theories, is presented. Extending previous results \ci{KorNic96} the
two-time Hamiltonian formulation is obtained, which describes the
complete phase space of the model in the isomonodromic sector. The
Dirac brackets arising from the coset constraints are
calculated. Their quantization allows to relate exact solutions of the
corresponding Wheeler-DeWitt equations to solutions of a modified
(Coset) Knizhnik-Zamolodchikov system.

On the classical level, a set of observables is identified, that is
complete for essential sectors of the theory. Quantum counterparts of
these observables and their algebraic
structure are investigated. Their status in alternative quantization
procedures is discussed, employing the link with Hamiltonian
Chern-Simons theory.\medskip

\end{abstract}
\hrule
\vspace*{1.6cm}
\section{Introduction}

It is an important class of physical theories, that admit the
formulation as a gravity coupled coset space $\s$-model after
dimensional reduction to two dimensions. Including pure gravity and
Kaluza-Klein theories as well as extended supergravity theories, in
$3\!+\!1$ dimensions they are described by a set of scalar and vector
fields coupled to gravity, where the scalar fields already form a
non-linear $\s$-model. Further reduction is achieved by imposing
additional symmetries --- manifest by assuming two additional
commuting Killing vector fields, for example corresponding to the
study of axisymmetric stationary models.

This reduction to effectively two dimensions leads to a non-linear
$\s$-model in an enlarged coset space, coupled to two-dimensional
gravity and a dilaton field. The arising additional scalar fields that
contribute to parametrizing the coset space are remnant of the
original vector fields and of components of the former
higher-dimensional metric. For general reason, related to boundedness
of the energy, it is the maximal compact subgroup $H$ of $G$ that is
divided out in the coset.  The first reduction of this type,
discovered for pure gravity \ci{Gero71}, leads to the simplest coset
space $\coset$. It was generalized up to the case of maximally
extended $N\!=\!8$ supergravity, where the $E_{8(+8)}/SO(16)$ arises
\ci{Juli80,Juli83}. The general proceeding was analyzed in
\ci{BrMaGi88,Nico91}.  \medskip

In \ci{KorNic95a,KorNic95b,KorNic96} a program was started
to perform an exact quantization of these dimensionally reduced
gravity models. Progress has been achieved using methods and techniques
similar to those developed in the theory of flat space integrable
systems \ci{Fadd84,FadTak87,KoBoIz93}. Despite the fact that
dimensional reduction via additional symmetries represents an
essential truncation of the theory, these so-called
``Midi-Superspace'' models under investigation are sufficiently
complicated to justify the hope, that their exact quantization might
provide insights into fundamental features of a still outstanding
quantized theory of gravitation. In particular and in contrast to
previously exactly quantized ``Mini-Superspace'' models, they
exhibit an infinite number of degrees of freedom, which is broadly
accepted to be a sine qua non for any significant model of quantum
gravity (compare \ci{Kuch71,AshPie96} for a discussion of this point in
the context of related models).

One of the final purposes of this approach is the identification of
exact quantum states, whose classical limit corresponds to the known
classical solutions.  For pure gravity this includes the quantum
analogue of the Kerr solution describing the rotating black hole; for
extended supergravities recently discovered corresponding solutions
have been of particular interest exhibiting fundamental duality
symmetries \ci{CveYou95,CveHul96}, such that their exact quantum
counterparts should shed further light onto the role of these
symmetries in a quantized theory.
\medskip

The main ideas of the new framework are the following: Exploiting the
integrability of the model, new fundamental variables have been
identified (certain components of the flat connection of the auxiliary
linear system continued into the plane of the spectral parameter), in
terms of which the ``right" and ``left" moving sectors have been
completely decoupled \ci{KorNic95a}.  The quantization is further
performed in the framework of a generalized ``two-time'' Hamiltonian
formalism, i.e.~these sectors are quantized independently.  The whole
procedure has been established in that sector of the theory, where the
new fundamental connection exhibits simple poles at fixed
singularities.
\medskip

In the present paper we achieve the consistent general formulation of
the desired coset-models in this approach. So far the formalism was
mainly elaborated in the technically simplified principal model, where
the coset $G/H$ had been replaced by the group $G$ itself. For the
coset model the phase space spanned by the new variables is too large
and must be restricted by proper constraints. Their canonical
treatment requires a Dirac procedure, which effectively reduces the
degrees of freedom. It leads to a consistent analogous Hamiltonian
formulation of the coset model allowing canonical quantization. Exact
quantum states are shown to be in correspondence to solutions of a
modified (Coset) Knizhnik-Zamolodchikov system.

Moreover, the formalism is kept general as long as possible, without
restricting to the simple pole sector. In particular, we completely
extend it to the case of connections with poles of arbitrary high
order at fixed singularities, which span the isomonodromic sector of
the theory. Generalization of the scheme to the full phase space
is sketched in Appendix \ref{beyi}.
\medskip

The other main result of this paper is the identification of classical
and quantum observables. For the above mentioned simple pole sector,
these sets are complete. Natural candidates for classical observables
are the monodromies of the fundamental connection in the plane of the
variable spectral parameter. We determine their (quadratic) Poisson
structure. After quantization of the connection quantum counterparts
of these monodromy matrices are identified as monodromies of certain
higher-dimensional KZ-systems.  Following Drinfeld \ci{Drin89a} their
algebraic structure may be determined to build some quasi-associative
braided bialgebra. The classical limit of this structure coincides
with the Poisson algebra of the classical monodromies found above. In
this sense, complete consistency of the picture is established.

The weakened coassociativity leads to a quantum algebra of observables
with operator-valued structure constants. This might have been avoided
by directly quantizing the regularized classical algebra of
monodromies, as is common in Chern-Simons theory
\ci{AlGrSc95a,AlGrSc95b}, instead of recovering quantum monodromies in
the picture of the quantized connection. We discuss this link and its
consequences.

The treatment of observables is performed in great detail for the
simplified principal model mentioned above. This is for the sake of
clarity of the presentation, since the arising difficulties in the
coset case deserve an extra study in the sequel. However, the main
tools and strategies that will finally be required can already and
clearer be developed and used in this context. The modifications
required for the coset model are clarified afterwards.
 \medskip

The paper is organized as follows. In Chapter 2 we start by
introducing the known linear system associated to the model and
describe the related on-shell conformal symmetry. A short summary and
generalization of the results from \ci{KorNic95a,KorNic96} about the
classical treatment of the principal model is given without
restricting to the simple pole sector. The link to Hamiltonian
Chern-Simons theory is discussed, where the same holomorphic Poisson
structure is obtained by symplectic reduction of the complexified
phase space in a holomorphic gauge fixing. This link in particular
enables us to relate the status of observables in both
theories. Observables in terms of monodromy matrices are identified;
their Poisson structure is calculated and discussed. The technical
part of the calculation is shifted into Appendix \ref{pmono}.

Chapter 3 treats the quantization of the principal model. We first
briefly repeat the quantization of the simple pole sector of this
model \ci{KorNic95b,KorNic96}. Quantum analogues of the
monodromy matrices are defined. Their algebraic structure and its
classical limit are determined and shown to be consistent with the
classical results. The alternative treatment in Chern-Simons theory
and the identification of quantum observables in these approaches are
discussed.

In Chapter 4 we finally present the generalization of the formalism to
the coset models.  A Hamiltonian formulation in terms of modified
fundamental variables is provided. The coset constraints are
explicitly solved by a Dirac procedure. Furthermore, we quantize the
simple pole sector of the coset model, showing that solutions of a
modified Knizhnik-Zamolodchikov-system identify physical quantum
states, i.e.~exact solutions of the Wheeler-DeWitt equations. We close
with a sketch of how to employ the whole machinery to the simplest
case of pure four-dimensional axisymmetric stationary gravity. In
particular, the existence of normalizable quantum states is shown.
Chapter 5 briefly summarizes the open problems for future work.

\boldmath
\section{Principal $\s$-model coupled to two-dimensional dilaton
gravity}
\unboldmath

The model to be studied in this paper is described by the following 
two-dimensional Lagrangian:
\be \label{Lagrangian}
\cL = e\r \Big(R + 
h^{\m\nu}\tr[\partial_\m gg^{-1}\partial_\nu gg^{-1}]\Big)
\ee
Here, $h_{\m\nu}$ is the $2D$ (``worldsheet'') metric, $e=\sqrt{|\det
h|}$, $R$ is the Gaussian curvature of $h^{\m\nu}$, $\r\in\R$ is the
dilaton field and $g$ takes values in some real coset space $G/H$,
where $H$ is the maximal compact subgroup of $G$. The currents
$\partial_\m gg^{-1}$ therefore live in a fixed faithful
representation of the algebra ${\mathfrak g}$ on some auxiliary
$d_0$-dimensional space $V_0$. It is well known that this type of
model arises from the dimensional reduction of higher dimensional
gravities \cite{BrMaGi88,Nico91}, e.g.~from $4D$ gravity in the
presence of two commuting Killing vectors \cite{BreMai87}. In the
latter case which describes axisymmetric stationary gravity, the
relevant symmetric space is $G/H = \coset$.

Let us first briefly describe further reduction of the Lagrangian
(\ref{Lagrangian}) by means of gauge fixing and state the resulting
equations of motion.  The residual freedom of coordinate
transformations can be used to achieve conformal gauge of the $2D$
metric $h_{\m\nu}$:
\ben
h_{\m\nu}{\rm d}x^{\m}{\rm d}x^{\nu} = h(z,\zb) {\rm d}z{\rm d}\zb,
\een
with world-sheet coordinates $z,\zb$, which reduces the Lagrangian to
\be
{\cal L} =\rho \Big( h R + {\rm tr} [g_{z} g^{-1}
g_{\zb}g^{-1}] \Big).
\la{L}
\ee
In this gauge the Gaussian curvature takes the form $R=(\log
h)_{z\bar{z}}/h$.  The equation of motion for $\rho$ derived from
\Ref{L}
\be
\rho_{z\zb} = 0 
\la{dilaton}
\ee
is solved by $\rho (z,\zb) = {\rm Im}
\, \x( z)$, where $\x(z)$ is a (locally) holomorphic function. 
  Then the equations of motion for $g$ coming
from \Ref{L} read
\be\la{ee1}
\big((\x-\xb)g_{z}g^{-1}\big)_{\zb} + 
\big((\x-\xb)g_{\zb}g^{-1}\big)_{z} =0 .
\ee
We can further specialize the gauge by identifying $\x,\xb$ with the
worldsheet coordinates. Then \Ref{ee1} turns into
\be\la{ee}
\big((\x-\xb)g_{\x}g^{-1}\big)_{\xb} + 
\big((\x-\xb)g_{\xb}g^{-1}\big)_{\x} =0 .
\ee

The equations of motion for the conformal factor are derived from the 
original Lagrangian (\ref{Lagrangian}):
\be\la{h}
(\log h)_{\x} =\f{\x-\xb}{4}\tr(g_{\x} g^{-1})^2 \quad{\rm and~c.c.}
\ee

Throughout this whole chapter we will for above mentioned reasons of
clarity investigate the simplified model, where the symmetric space
$G/H$ is replaced by the group $G$ itself. We will refer to this
plainer model as the {\bf principal model}.

\subsection{Linear system and on-shell conformal symmetry of the model}

The starting point of our treatment is the following well-known linear
system associated to the equations (\ref{ee}) \ci{BelZak78,Mais78}:
\be
\f{d\Psi}{d\x}= \f{g_{\xi}g^{-1}}{1-\g}\Psi,\qquad
\f{d\Psi}{d\xb}= \f{g_{\xb}g^{-1}}{1+\g}\Psi,
\la{ls}\ee
where $\g$ is the spacetime-coordinates dependent ``variable spectral
parameter'':
\be
\g=\f{2}{\x-\xb}\left\{w -\f{\x+\xb}{2} 
\pm\sqrt{(w-\x)(w-\xb)}\right\}
\la{gamma}
\ee
or alternatively $w\in \C$ may be interpreted as a hidden ``constant
spectral parameter''; $\Psi(w,\x,\xb)$ is a $G_\C$-valued
function. The variable spectral parameter $\g$ lives on the twofold
covering of the complex $w$-plane, the transition between the sheets
being performed by $\g\!\mapsto\!\frac1{\g}$. It satisfies:
\be
\f{\p \g}{\p\x}=\f{\g}{\x-\xb}\f{1+\g}{1-\g}, 
\qquad
\f{\p\g}{\p\xb}=\f{\g}{\xb-\x}\f{1-\g}{1+\g} 
\la{pe}
\ee
such that in \Ref{ls} it is:
\be\la{der}
\f{d}{d\x}=\f{\p}{\p\x}+\f{\g}{\x-\xb}\f{1+\g}{1-\g}\f{\p}{\p\g},
\qquad
\f{d}{d\xb}=\f{\p}{\p\xb}+\f{\g}{\xb-\x}\f{1-\g}{1+\g}\f{\p}{\p\g}
\ee

The linear system \Ref{ls} exists due to the following on-shell
M\"obius symmetry of equations of motion.\footnote{A similar symmetry
exists in the theory of Bianchi surfaces \ci{Bobe93}.}

\begin{Theorem}
Let $g(z,\zb)$, $\rho(z,\zb)= {\rm Im} \xi(z)$ and $h(z,\zb)$ be some
solution of \Ref{dilaton}, \Ref{ee1}, \Ref{h} and $\Psi$ be the
related solution of the linear system \Ref{ls}. Then
\be
\s^w[g] \equiv \Psi^{-1}\left(\f{1}{\g}\right)\Psi(\g)\,,\qquad
\s^w[\x] \equiv\f{w \x(z)}{w-\x(z)}\,,\qquad
\s^w[h] \equiv h \la{Mosy}
\ee
also solve \Ref{ee1},\Ref{h}
\end{Theorem}
\paragraph{Proof:} 
We have
\ba
\s^w[ g_\x g^{-1}]&=&\sqrt{\f{w-\xb}{w-\x}}
\Psi^{-1}\left(\f{1}{\g}\right) g_\x g^{-1}
\Psi\left(\f{1}{\g}\right)\non
\s^w[ g_\xb g^{-1}]&=&\sqrt{\f{w-\x}{w-\xb}}
\Psi^{-1}\left(\f{1}{\g}\right) g_\xb g^{-1}
\Psi\left(\f{1}{\g}\right)\nn
\ea
Now fulfillment of \Ref{ee1},\Ref{h} may be checked by straightforward
calculation.
\qed

The transformations $\s^w$ form a one-parametric abelian subgroup of 
the group $SL(2,\R)$ of conformal transformations. We have
\ben
\s^{w_1} \s^{w_2} = \s^{w_3}\,,\qquad \f{1}{w_1} +\f{1}{w_2}= 
\f{1}{w_3}
\een
The full M\"obius group may be obtained combining transformations
$\s^w$ with (essentially trivial) transformations
\ben
\xi(z)\mapsto a\xi(z) + b\,,\qquad
g(z)\mapsto g(z)
\een
which obviously leave the equations of motion invariant.  As a result
the action of an arbitrary $SL(2,\R)$ M\"obius transformation $\s$ on
the solution of equations of motion is
\be\la{cs}
\xi(z) \mapsto\s[\x]\equiv a\f{w \x(z)}{w-\x(z)} + b\,,\qquad
g(z,\bar{z})\mapsto\s[g]\equiv\Psi^{-1}\left(\f{1}{\g}\right)\Psi(\g)\,,
\ee
leaving $h$ invariant.  In addition to the M\"obius symmetry \Ref{cs}
the model possesses the symmetry corresponding to an arbitrary
holomorphic change of the worldsheet coordinate $z$ (this symmetry
disappears if we identify $z$ with $\x$).  Combining this symmetry
with \Ref{cs} reveals the following M\"obius symmetry of equation
\Ref{ee}
\ba
g(\x,\xb) &\mapsto& \s[g] 
\left(\f{w(\x-b)}{aw+\x-b},\f{w(\xb-b)}{aw+\xb-b}\right)\la{cs1}\\
h(\x,\xb) &\mapsto& h 
\left(\f{w(\x-b)}{aw+\x-b},\f{w(\xb-b)}{aw+\xb-b}\right)
\la{cs2}\ea
Infinitesimally, the symmetry \Ref{cs1} is a subalgebra of the
Virasoro symmetry of \Ref{ee} \ci{JulNic96}.

\begin{Note}\rm
It is known that the Ernst equation \Ref{ee1} for $SL(2,\R)/SO(2)$ may
be rewritten as a fourth order differential equation in terms of the
conformal factor $h$. The transformation \Ref{cs2} shows that this
equation is, in contrast to Ernst equation itself, M\"obius invariant
in the $\x, \xb$-plane.
\end{Note}

\subsection{Two-time Hamiltonian formulation of the principal model}
Here we present a generalized version of the ``two-time'' Hamiltonian
formalism of the principal $\s$-model proposed in
\ci{KorNic95a,KorNic95b}. It is the strategy to define a new set of
fundamental variables by means of exploiting the corresponding linear
system. These variables may be equipped with a Poisson structure such
that a two-time Hamiltonian formulation of the model is
achieved.

\subsubsection{New fundamental variables and the
isomonodromic sector}

The main objects we are going to consider as fundamental variables in
the sequel are certain components of the following one-form
\begin{Definition}
Let $\Psi(\g,\x,\xb)$ be a solution of the linear system
(\ref{ls}). Then the ${\mathfrak g}$-valued one-form $\bA$ is defined
as:
\be\la{A}
\bA := {\rm d}\Psi\Psi^{-1}
\ee
In particular, we are interested in the components
\be\la{AgAw}
\bA = A^{\g}{\rm d}\g + A^{\x}{\rm d}\x + A^{\xb}{\rm d}\xb
= A^{w}{\rm d}w + \tilde{A}^{\x}{\rm d}\x + \tilde{A}^{\xb}{\rm d}\xb
\ee
where $(\g,\x,\xb)$ and $(w,\x,\xb)$ respectively are considered to 
be independent variables. In the sequel we shall use the shortened 
notation $A\equiv A^{\g}$.

Moreover, we will restrict our study to that sector of the theory,
where $A$ is a single-valued meromorphic function of $\g$, i.e.~that
also $\bA$ is single-valued and meromorphic in $\g$. A solution $\Psi$
of \Ref{ls} with this property is called {\bf isomonodromic}, as its
monodromies in the $\g$-plane then have no $w$-dependence due to
\Ref{A}.
\end{Definition}

Further on, we immediately get the following relations:
\begin{Lemma}\la{Lcur}
The relation of the original field $g$ to $A$ is given by
\be\la{cur}
g_{\x} g^{-1}=\left.\f{2}{\x-\xb}\, A(\g,\x,\xb)\right|_{\g=1} \qquad
g_{\xb} g^{-1}=\left.\f{2}{\x-\xb}\, A(\g,\x,\xb)\right|_{\g=-1}
\ee
as a corollary of \Ref{ls} and \Ref{der}. Moreover, the linear system
(\ref{ls}) and definition (\ref{AgAw}) imply:
\ba
A^w &=& \frac{\p\g}{\p w} A\\
\tilde{A}^{\x} &=& \frac{2A(1)}{(\x-\xb)(1-\g)}, \qquad
\tilde{A}^{\xb} ~=~ \frac{2A(-1)}{(\x-\xb)(1+\g)} \non
A^{\x} &=& \f{2A(1)-\g(1+\g)A(\g)}{(\x-\xb)(1-\g)}, \qquad
A^{\xb} ~=~ \f{2A(-1)+\g(1-\g)A(\g)}{(\x-\xb)(1+\g)} \nn
\ea
\end{Lemma}
\qed

\begin{Note}\rm
In the sequel $A(\g)$ will be exploited as the basic fundamental
variable. At this point we should stress the difference between the
real group $G$ (with algebra ${\mathfrak g}$) entering the physical
models and the related complexified group $G_\C$ (with algebra
${\mathfrak g}_\C$). Namely, it is $A(\g\!\in\!\C)\in {\mathfrak
g}_C$, whereas we will additionally impose the ``imaginary cut''
$iA(\g\!\in\!i\R)\in {\mathfrak g}$. Since $A(\g)$ is a (locally)
holomorphic function, this implies
\be\la{realcon}
A(\bar{\g})=- A^*(-\g)
\ee
where $^*$ denotes the anti-linear conjugation on ${\mathfrak g}_\C$
defined by the real form ${\mathfrak g}$. Together with \Ref{cur} this
ensures $g\in G$.  
\end{Note}

\begin{Note}\rm
The linear system \Ref{ls} admits the normalization
\be\la{norma}
\Psi(\g\!=\!\infty)=I,
\ee
which implies regularity of $A$ at infinity:
\be\la{cona}
A_\i:=\lim_{\g\rightarrow\infty}\g A(\g) = 0
\ee
Furthermore, \Ref{ls} implies an additional relation between the
original field $g$ and the $\Psi$-function:
\be\la{C0}
\Psi(\g\!=\!0) = gC_0,
\ee
where $C_0$ is a constant matrix in the isomonodromic sector.
\end{Note}
\bigskip

The definition of $\bA$ as pure gauge (\ref{A}) implies integrability
conditions on its components, which in particular give rise to the
following closed system for $A(\g)$:
\be\la{A1}
\f{\p A}{\p\x}= [A^{\x}, A] + \f{\p A^{\x}}{\p\g},\qquad
\f{\p A}{\p\xb}= [A^{\xb}, A] + \f{\p A^{\xb}}{\p\g}
\ee

The main advantage of the system \Ref{A1} in comparison with the
original equations of motion in terms of $g$ \Ref{ee} is, that the
dependence on $\x$ and $\xb$ is now completely decoupled.  Once the
system \Ref{A1} is solved, it is easy to check that the equations
\Ref{cur} are compatible and the field $g$ restored by means of them
satisfies \Ref{ee}.

The remaining set of equations of the principal model \Ref{h}, which
concern the conformal factor $h$, may be rewritten taking into
account \Ref{cur} as the following constraints:
\be\la{C}
\Cx:= 
- (\log h)_{\x} +\f{1}{\x-\xb}\tr A^2(1)=0,\qquad
\Cxb:= 
- (\log h)_{\xb} +\f{1}{\xb-\x}\tr A^2(-1)=0
\ee

\subsubsection{Poisson structure and Hamiltonians}
The described decoupling of $\x$ and $\xb$ dependence allows to treat
the system \Ref{A1}, \Ref{C} in the framework of a manifestly
covariant two-time Hamiltonian formalism, where the field $A(\g)$, the
``times'' $\x$, $\xb$ and the fields $(\log h)_\x$, $(\log h)_\xb$ are
considered as new basic variables. The spirit of generalized
``several-times'' Hamiltonian formalism is described for example in
\ci{Kast83,Dick91}.

For this purpose we equip $A(\g)$ with the following
(equal $\x, \xb$) Poisson structure:
\begin{Definition}
Define the Poisson bracket on $A(\g)\equiv A^a(\g)t_a$ as:
\be
\left\{A^a(\g)\,,\,A^b(\m)\right\} =
-f^{abc}\frac{A^c(\g)-A^c(\m)}{\g-\m},
\la{PB1}
\ee
$f^{abc}$ being the structure constants of ${\mathfrak
g}$.\footnote{Assuming ${\mathfrak g}$ to be semisimple, the existence
of the symmetric Killing-form enables us to arbitrarily pull up and
down the algebra indices.}
\end{Definition}

The relations
\ba
\left\{A(\g)\,,\,\frac1{\x-\xb}\tr A^2(1)\right\} 
&=& \left[A^\x(\g)\,,\,A(\g)\right], \la{imdyn}\\
\left\{A(\g)\,,\,\frac1{\xb-\x}\tr A^2(-1)\right\} 
&=& \left[A^\xb(\g)\,,\,A(\g)\right],\nn
\ea
compared with the equations of motion (\ref{A1}) give rise to
\begin{Definition}\la{impexp}
We call the ($\x, \xb$)-dynamics that is generated by 
\be\la{H}
H^\x :=\frac1{\x-\xb}\,\tr A^2(1) ,\qquad
H^\xb :=\frac1{\xb-\x}\,\tr A^2(-1) ,
\ee
the {\bf implicit time dependence} of the fields. The remaining ($\x,
\xb$)-dynamics is referred to as {\bf explicit time dependence}.
\end{Definition} 

In fact, the motivation for this definition arises from
\ci{KorNic95a,KorNic95b}, where it has been shown, that in essential
sectors of the theory (simple pole singularities in the connection
$A$), it is possible to identify a complete set of explicitly
time-independent variables. They may be treated as canonical variables
then, such that $H^\x$ and $H^\xb$ serve as complete
Hamiltonians. This will be illustrated and generalized in the next
subsections for the isomonodromic sector of the theory, where $A(\g)$
is assumed to be a meromorphic function of $\g$. 

The extension of this framework to the whole phase space of arbitrary
connections $A$, that is strongly inspired from the treatment of the
simple pole case, is sketched in Appendix \ref{beyi}.  The variables
$A(\g)$ themselves are explicitly time-dependent in general according
to \Ref{A1} and
\Ref{imdyn}.

\begin{Note}\rm
The set of 
\be
B(w) = A^w (\g)+ A^w \left(\f{1}{\g}\right) 
\equiv \f{\p\g}{\p w}\left(A(\g)-
\f{1}{\g^2}A\left(\f{1}{\g}\right)\right)
\la{B}\ee
build a rather simple set of explicitly time-independent variables,
carrying half of the degrees of freedom of the full phase space.
This may be checked by straightforward calculation. 
Moreover, \Ref{PB1} implies
\be
\{B^a(w),B^b(v)\} = -f^{abc}\frac{B^c(w)-B^c(v)}{w-v}
\la{PBB1}
\ee
\end{Note}

\begin{Note}\la{comPB}\rm
From the mathematical point of view, \Ref{PB1} is a rather natural
structure \ci{FadTak87}, even though it is not canonically derived from
the Lagrangian \Ref{Lagrangian}. It may however be obtained from an
alternative Chern-Simons Lagrangian formulation of the model, as is
sketched in the following section. Comparison to the conventional
Poisson structure of \Ref{Lagrangian} should be worked out on the
space of observables, where due to spacetime-diffeomorphism invariance
no principal difference between one- and two-time structures appears.
\end{Note}

In order to gain a Hamiltonian description for the total ($\x,
\xb$)-dependence of the fields, we employ a full covariant treatment
by additionally introducing conjugate momenta for the canonical
``time'' variables $\x$ and $\xb$.
\begin{Definition}
Define the (equal $\x, \xb$) Poisson bracket
\be\la{PB2}
\Big\{ \x, -(\log h)_\x \Big\} = \Big\{ \xb, -(\log h)_\xb
\Big\} = 1,
\ee
where in the sense of a covariant theory only the explicit appearance
of $\x, \xb$ (compare Def. \ref{impexp}) is covered by treating these
previous ``times'' as additional canonical variables, which obey the
bracket \Ref{PB2}.
\end{Definition}

This identification of the conjugate momenta for the explicitly
appearing times with the logarithmic derivatives of the conformal
factor is motivated from the Lagrangian (\ref{L}) \ci{NiKoSa96}. It
implies that the dynamics in $\x$ and $\xb$ directions is completely
given by the Hamiltonian constraints $\Cx$ and $\Cxb$ defined in
(\ref{C}), i.e.~for any functional $F$ we have
\be
\f{d F}{d\x} = \{ F,\Cx \}, \qquad 
\f{d F}{d\xb} = \{F,\Cxb \}.  \la{evolve}
\ee
The remaining equations of motion \Ref{C} mean weak vanishing of the
Hamiltonians. This phenomena always arises in the framework of
covariant Hamiltonian formalism when time is treated as canonical
variable in its own right canonically conjugated to the Hamiltonian
\ci{HenTei92}; it is a standard way to take into account possible
reparametrization of the time variable.

\subsubsection{First order poles}\la{fop}

In this simplest case considered in \ci{KorNic95a,KorNic96} we assume
that $A(\g)$ has only simple poles, i.e.
\be
A(\g)=\sum_{j=1}^N \f{A_j(\x,\xb)}{\g-\g_j},
\la{simplep}\ee
where according to \Ref{ls} all $\g_j$ should satisfy \Ref{pe},
i.e.~$\g_j=\g(w_j,\x,\xb)$, $w_j\in \C$.  Then the equations of motion
\Ref{A1} yield
\be
\f{\p A_j }{ \p \x } =
\f{2}{\x-\xb}\sum_{k\neq j}\f{[A_k,\;A_j]}{(1-\g_k)(1-\g_j)}
,\qquad
\f{\p A_j }{ \p \xb } =
\f{2}{\xb-\x}\sum_{k\neq j}\f{[A_k,\;A_j]}{(1+\g_k)(1+\g_j)}
\la{1}
\ee 
and the Poisson brackets
\Ref{PB1} and \Ref{PB2} reduce to
\ba
\{A_i^a,A_j^b\} &=& \d_{ij}f^{abc}A_j, \la{PBAsp}\\
\{A_j,(\log h)_\x\}&=&\{A_j,(\log h)_\xb\} ~=~0, \non
\{\g_j,(\log h)_\x\} &=& -\p_\x \g_j, \\
\{\g_j,(\log h)_\xb\} &=& -\p_\xb \g_j, \nn
\ea
i.e.~in this case, the residues $A_j$ together with the set of (hidden
constant) positions of the singularities $\{w_j\}$ give the full set
of explicitly time-independent variables.

\subsubsection{Higher order poles}\la{hp}

We can also generalize the described formulation to the case, where
$A(\g)$ has higher order poles in the $\g$-plane:

\be\la{Ahp}
A(\g)=\sum_{j=1}^N\sum_{k=1}^{r_j} \f{A_j^k(\x,\xb)}{(\g-\g_j)^k}
\ee

The Poisson structure \Ref{PB1} in terms of $A_j^k$ has the following
form:
\be\la{PBAh}
\{(A_i^k)^a, (A_{j}^{l})^b\}  = \cases{
  \begin{array}{ll} \delta_{ij} f^{abc} (A_j^{k+l-1})^c   
       & \mbox{for $k+l-1\leq r_j$} \\ 0 & \mbox{for $k+l-1> r_j$}
  \end{array} },
\ee
building a set of mutually commuting truncated half affine algebras.

However, it turns out that for $r_j>1$ the variables $A_j^k$ for
$k\!=\!1,\dots r_j\!-\!1$ have non-trivial Poisson brackets with
$(\log h)_\x$ and $(\log h)_\xb$, and, therefore, are not explicitly
time-independent.
The problem of identification of explicitly time-independent
variables can be solved in the following way. Consider
\ben
A^w(\g)= \f{\p\g}{\p w} A(\g)
\een
which as a function of $w$ is meromorphic on the twofold covering of
the $w$-plane. Parametrize the local expansion of $A^w$ around one of
its singularities $\g_j$ as:
\be\la{exphp}
A^w(\g)=\sum_{k=1}^{r_j} \f{A_j^{(w)k}}{(w-w_j)^k} + \cO((w-w_j)^0)
\qquad{\rm for}\quad\g\sim \g_j
\ee

We can now formulate
\begin{Theorem}
The coefficients $A_j^{(w)k}$ of the local expansion of $A^w$ have no
explicit time dependence, i.e.
\be\la{Awdyn}
\p_\x A_j^{(w)k} = \{A_j^{(w)k},H^\x\},\qquad
\p_\xb A_j^{(w)k} = \{A_j^{(w)k},H^\xb\}.
\ee
They satisfy the same Poisson structure as the $A_j^k$ \Ref{PBAh}:
\be\la{PBw2}
\left\{(A_i^{(w)k})^a\,,\, (A_{j}^{(w)l})^b\right\}  = \cases{
  \begin{array}{ll} \delta_{ij} f^{abc} (A_j^{(w)k+l-1})^c   
       & \mbox{for $k+l-1\leq r_j$} \\ 0 & \mbox{for $k+l-1> r_j$}
  \end{array} }
\ee
\end{Theorem}
\paragraph{Proof:}
Let us first prove \Ref{Awdyn}. From \Ref{PB1} and the definition of
$H^\x$ it follows that:
\ben
\{A^w(\g),H^\x\} ~=~ \{\p_w\g A(\g)~,~\frac{2\tr A^2(1)}{(\x-\xb)} \} 
~=~ \frac{\p_w\g}{(\x-\xb)} \left[\frac{2A(1)}{1-\g},A(\g)\right] 
~=~ [\tilde{A}^\x(\g),A^w(\g)] 
\een
whereas from \Ref{A} the $\x$-dynamics of $A^w$ is determined to be:
\ben
\p_\x A^w ~=~ [\tilde{A}^\x(\g),A^w(\g)] + \p_w \tilde{A}^\x(\g) 
~=~ [\tilde{A}^\x(\g),A^w(\g)] + \p_w\g\frac{2A(1)}{(1-\g)^2} 
\een

As the last term is regular in $\g\!=\!\g_j$, comparison of the two
previous lines shows that the $\x$-dependence of the coefficients in
the $w$-expansion around these points is completely generated by
$H^\x$, which proves \Ref{Awdyn}.

To show the Poisson structure \Ref{PBw2}, one has to consider the
corresponding coefficients of singularities in \Ref{PB1}. For
$i\!\not=\!j$, the result follows directly from \Ref{PBAh}, as
$A_j^{(w)k}$ is a function of $A_j^l, l\!=\!1,\dots,r_j$ only, such
that locality remains. For $i\!=\!j$, one may first extract from
\Ref{PB1} the behavior of $\{A^w(\g),A^w(\m)\}$ around $\g\sim\g_j$:
\ben
\{(A^w)^a(\g),(A^w)^b(\m)\} ~=~ 
-\p_w\g\p_v\m f^{abc}\frac{A^c(\g)-A^c(\m)}{\g-\m} ~\sim~
f^{abc} \frac{(A^w)^c(\g)}{\m-\g} \p_v\m
\een
to then further study the asymptotical behavior $\m\sim\g$:
\ben
\{(A^w)^a(\g),(A^w)^b(\m)\} ~\sim~ f^{abc} \frac{(A^w)^c(\g)}{v-w}   
\een
such that \Ref{PBw2} for $i\!=\!j$ follows in the same way, as does
\Ref{PBAh} from \Ref{PB1}.
\qed

Thus, also in this case we have succeeded in identifying a
complete set of canonical explicitly time-independent variables.

\begin{Note}\rm
Comparing \Ref{Ahp} with \Ref{exphp} shows, that the $A_j^{(w)k}$ are
related to the $A_j^l$ by means of explicit recurrent relations that
may be derived, expanding \Ref{Ahp} in $(w\!-\!w_j)$. Then
$A_j^{(w)k}$ is a function of $A_j^l$ with $k\leq l\leq r_j$. In
particular, the residues of highest order are related by:
\ben
\left(\frac{\p\g_j}{\p w_j}\right)^{r_j-1} A_j^{(w)r_j} = A_j^{r_j},
\een
which explains for example, why this difference was not relevant in the
case of simple poles in the last subsection.
\end{Note}

\subsection{The link to Hamiltonian Chern-Simons theory}
The treatment of the principal model of dimensionally reduced gravity
in the previous section was inspired by the fact, that the equations
of motion were obtained as compatibility conditions (\ref{A1}) of
special linear systems. The interpretation of these equations as zero
curvature conditions suggests a link with Chern-Simons theory whose
equations of motion also state the vanishing of some curvature. The
Chern-Simons gauge connection then lives on a space locally
parametrized simultaneously by the spectral parameter $\g$ and one of
the true space time coordinates playing the role of time.

The relevant Chern-Simons action reads
\be\label{CSaction}
S = \frac{k}{4\pi}\int_M \tr[\bA{\rm d}\bA - \frac23\bA^3],
\ee
where $\bA$ is a connection on a trivial $G$ principal bundle over the
3-dimensional manifold $M$. In the case of interest here, the manifold
$M$ is the direct product of the Riemann surface $\S$, on which the
spectral parameter $\g$ lives, and the real axis, which is
interpreted as time. For this configuration, Chern-Simons theory is
known to have a Hamiltonian formulation. Choosing proper boundary
conditions on the connection, the action may be rewritten in the form

\be\label{CSaction2}
S=-\frac{k}{4\pi}\int_M \tr[A\partial_tA] {\rm d}t + 
\frac{k}{2\pi}\int_M \tr[A^0({\rm d}A-A^2)]{\rm d}t 
\ee

The connection has been split $\bA=A+A^0{\rm d}t$ into spatial and time
components, where $A^0$ now plays the role of a Lagrangian multiplier
for the constraint
\be
F = {\rm d}A - A^2 = 0 \label{vc} 
\ee

Usually, $A^0$ is gauged to zero which leads to static components
$A$. In particular, any singularities of the connection are
time-independent in this case and treated by inserting static Wilson
lines in the action (\ref{CSaction2}) \cite{Witt89,ElMoScSe89}. A
nontrivial and somewhat singular gauge for $A^0$ must be chosen, to
derive the equations of motion of the described principal model of
dimensionally reduced gravity.

The further required holomorphic reduction of Chern-Simons theory can
still be described for arbitrary gauge fixing of $A^0$, as the results
will be valid in any gauge.

\subsubsection{Holomorphic reduction and Poisson bracket of the
connection}

For the following we first complexify the phase space and thereby also
the gauge group. This enlarged gauge freedom may be used for
a holomorphic gauge fixing then.

Denoting the spatial coordinates which locally parametrize $\S$ by
$\g\!=\!x+iy, \bar{\g}\!=\!x-iy$, defining the measure as
$\frac{k}{4\pi}{\rm d}x{\rm d}y \equiv \frac{-2i\k}{4\pi}{\rm d}x{\rm
d}y = \frac{\k}{4\pi}{\rm d}\g{\rm d}\gb$ and splitting the remaining
dynamical parts of $\bA$ into $A=A^{\g}{\rm d}\g+A^{\bar{\g}}{\rm
d}\bar{\g}$, the action (\ref{CSaction2}) implies the Poisson
structure:
\be
\{A^{\g,a}(\g,\gb),A^{\bar{\g},b}(\m,\mb)\} = 
-\frac{i\pi}{\k}\d^{ab}\delta^{(2)}(\g-\m), 
\label{ABbracket}
\ee
where here and in the following the $\delta$-function is understood as
a real two-dimensional $\delta$-function: 
$\delta^{(2)}(x+iy)\equiv\frac{i}2\delta(x)\delta(y)$, normalized such
that $\int {\rm d}\g{\rm d}\bar{\g}\delta^{(2)}(\g) = 1$.
\medskip

This Poisson structure corresponds to the Atiyah-Bott symplectic
form on the space of smooth connections on the Riemann surface $\S$
\cite{AtiBot82}:
\ben
\O = \frac{k}{4\pi}\rm{tr}\int_{\S} \delta A \wedge \delta A,
\een

The flatness constraints (\ref{vc}) are of the first class with
respect to this bracket:
\ben
\{F^a(\g,\gb),F^b(\m,\mb)\} = 
\frac{i\pi}{\k}f^{abc}F^c(\g)\delta^{(2)}(\g-\m),
\een
where $f^{abc}$ are the total antisymmetric structure constants of
${\mathfrak g}_\C$. These constraints generate the canonical gauge
transformations
\be
A \mapsto gAg^{-1} + dg g^{-1} \label{gt}
\ee
which leave the symplectic structure invariant.\medskip

The phase space of the original theory is therefore reduced to the
space of flat connections $A(\g,\gb)$ modulo the action of the complex
gauge group (\ref{gt}). If the singularities of the connection $A$ are
restricted to simple poles, this phase space is for instance
completely described by the monodromies of the connection. As a first
step to explicitly reduce the number of degrees of freedom, we will
fix the gauge freedom (\ref{gt}) in $A$, by demanding
\be 
A^{\bar{\g}}=0,\label{gf} 
\ee 
which makes flatness of $A(\g,\gb)$ turn into holomorphy of the
surviving component $A^{\g}(\g)$.

\begin{Note}\rm
The existence of corresponding gauge transformations is a nontrivial
problem. In general, when $A^{\bar{\g}}$ is gauged away, $A^{\g} {\rm
d}\g$ becomes a connection on a nontrivial bundle over $\S$. On
Riemann surfaces of higher genus, this form of gauge generically leads
to multivalued holomorphic quantities exhibiting certain twist
properties \cite{KorSam95}. On the Riemann sphere the gauge
transformations preserving single-valuedness of $A^{\g}{\rm d}\g$ at
least exist on a dense subspace of connections
\cite{AtiBot82,GawKup91}. For the purpose here, strictly speaking we a
priori restrict the phase space to the class of functions on the
punctured sphere that allow this gauge fixing. This includes e.g.~all
the connections with the curvature exhibiting $\d$-function
singularities treated in \ci{ElMoScSe89} (gauge fixed to holomorphic
connections with simple poles) as well as connections with higher
order derivatives of $\d$-functions in the curvature.
\end{Note}

This gauge fixing of first-class constraints changes the Poisson
structure according to Dirac \cite{Dira67}, leading to

\begin{Theorem}
Let the Poisson structure (\ref{ABbracket}) for the connection
\ben
A(\g,\gb) \equiv A^{\g,a}(\g,\gb)t_a{\rm d}\g +
A^{\gb,a}(\g,\gb)t_a{\rm d}\gb 
\een
be restricted by the constraints
(\ref{vc}) and (\ref{gf}). Then the Dirac bracket for the surviving
holomorphic components $A^a(\g) \equiv A^{\g,a}(\g)$ is given by:
\be
\{A^a(\g),A^b(\m)\}^* = 
\frac1{2\k}f^{abc}\frac{A^c(\g)-A^c(\m)}{\g-\m} \label{hbracket}
\ee
\end{Theorem}

\paragraph{Proof:}
The bracket between the constraints and
the gauge-fixing condition is of the form:
\be
\{F^a(\g),A^{\bar{\g},b}(\m)\} = 
\frac{i\pi}{\k}\delta^{ab}\partial_{\bar{\g}}\delta^{(2)}
(\g-\m) 
+\frac{i\pi}{\k} f^{abc}A^{\bar{\g},c}(\g)\delta^{(2)}
(\g-\m) \label{brcon}
\ee

On the constraint surface \Ref{gf} this matrix can be inverted using
$\partial_{\bar{\g}}\frac1{\g}\!=\!-2\pi i\delta^{(2)}(\g)$, which
follows from the inhomogeneous Cauchy theorem. The Dirac bracket for
the remaining holomorphic variables $A^{\g}(\g)$ then is:

\ben
\{A^{\g,a}(\g),A^{\g,b}(\m)\}^*
\een
\vspace*{-0.3cm}
\ba
\hspace*{1.5em}&=& - \sum_{m,n} 
\int {\rm d}x{\rm d}\bar{x}{\rm d}y{\rm d}\bar{y}\non
&&\left(
\{A^{\g,a}(\g),F^m(x)\}\left(\{F^m(x),A^{\bar{\g},n}(y)\}\right)^{-1}
\{A^{\bar{\g},n}(y),A^{\g,b}(\m)\}\right. \non
&&{} \hspace{-0.8em} + \left. \{A^{\g,a}(\g),A^{\bar{\g},n}(y)\}
\left(\{A^{\bar{\g},n}(y),F^m(x)\}\right)^{-1}\{F^m(x),A^{\g,b}(\m)\} 
\right) \non
\vspace*{2mm}
&=& -\sum_m\int {\rm d}x{\rm d}\bar{x}{\rm d}y{\rm d}\bar{y}
\frac{i\pi}{\k}\non
&&\left( 
\left(\d^{am}\partial_x\d^{(2)}(x-\g)+f^{mac}A^{\g,c}(x)\d^{(2)}(x-\g)
\right)\frac{\d^{mb}\d^{(2)}(y-\m)}{2\pi i(x-y)}\right.\non
&&{} \hspace{-0.6em} - \left.\left(\d^{bm}\partial_x\d^{(2)}(x-\m)+
f^{mbc}A^{\g,c}(x)\d^{(2)}(x-\m)\right)\frac{\d^{am}\d^{(2)}
(\g-y)}{2\pi i(x-y)}\right) \non
\vspace*{2mm}
&=& \frac1{2\k}f^{abc}\frac{A^{\g,c}(\g)-A^{\g,c}(\m)}{\g-\m} \nn
\ea
\qed

The holomorphic structure has in this context first been proposed by
Fock and Rosly \cite{FocRos92}.

\begin{Note}\rm
For convenience in concrete calculations we still give this result in
tensor notation, as is explicitly explained in \cite{FadTak87}, where
the relation of \Ref{hbracket} to the corresponding current algebra is
discussed. This structure may be put into the form
\be\la{hbracketR} 
\{A(\g)\stackrel{\otimes}{,}A(\m)\} = [r(\g-\m),A(\g)\otimes I +
I\otimes A(\m)] 
\ee 
with the classical $r$-matrix $r(\g)=-\frac1{2\k}\frac{\O}{\g}$, where
$\O=t^a\otimes t_a$ is represented as $d_0^2\times d_0^2$ matrix
here. For the simplest but important case ${\mathfrak
g}$=$\mathfrak{sl}(2)$, it is $\O=\frac12I\!\otimes\!I+\Pi$, with
$\Pi$ being the $4\times 4$ permutation operator. The matrix $r(\g)$
satisfies the classical Yang-Baxter equation with spectral parameter:
\be
[r^{12}(\g-\mu), r^{13}(\g)+ r^{23}(\mu)] + [r^{13}(\g),\; r^{23}(\mu)] =0
\la{CLYB}
\ee

In shortened notation, \Ref{hbracketR} reads:
\be
\{A(\g)^0,A(\m)^{\bar{0}}\} = 
[r(\g-\m),A(\g)^0 + A(\m)^{\bar{0}}], 
\la{PBR}
\ee 
with $A(\g)^0:=A(\g)\otimes I,\;\; A(\m)^{\bar{0}}:=I\otimes A(\m)$. 
\end{Note}

\begin{Note}\rm
In the framework of canonical and geometric quantization of
Chern-Simons theory \cite{Witt89,AxWiDe91,ElMoScSe89,GawKup91}, the
variables $A^{\g}$ and $A^{\bar{\g}}$ are --- according to
(\ref{ABbracket}) --- considered and treated as canonically conjugated
coordinate and momentum, respectively.  After the holomorphic gauge
fixing the surviving variable $A(\g)= A^{\g}(\g)$ resembles ---
according to
(\ref{hbracket}) --- a combination of angular momenta.
\end{Note}

\begin{Note}\rm
The flatness constraints (\ref{vc}) have not been totally fixed by the
choice of gauge (\ref{gf}). Apparently this gauge still admits
holomorphic gauge transformations, which on the sphere reduce to
constant gauge transformations. This freedom may also be seen from
the appearance of $\partial_{\bar{\g}}$ in the matrix of constraint
brackets (\ref{brcon}), which actually prevents its strict
invertibility.  This implies the surviving of the (global) first-class
part of the flatness constraint $F$, which for meromorphic $A$ in the
parametrization (\ref{Ahp}) is:
\be\label{con}
\int F^a(\g) {\rm d}\g{\rm d}\gb = 
\int \partial_{\bar{\g}} A^a(\g){\rm d}\g{\rm d}\gb 
= -2\pi i \sum_i (A_i^1)^a = -2\pi i A_\i^a, 
\ee
where $A_\i=A_\i^a t_a$, compare \Ref{cona}. Obviously, $A_\i^a$ is a
generator of constant gauge transformations in the bracket
(\ref{hbracket}).
\end{Note}

\subsubsection{Embedding the principal model}\la{embedd}
In this holomorphic structure of Chern-Simons theory the link to the
principal model can be established. As a first fact, note that the
Dirac bracket (\ref{hbracket}) for $\k\!=\!-\frac12$ equals the Poisson
structure (\ref{PB1}) that was used for the Hamiltonian formulation of
the principal model.

The equations of motion from Chern-Simons action (\ref{CSaction}) read
\be
\partial_t A^{\g} = \partial_{\g}A^0 + [A^{\g},A^0],
\ee
leading to trivial dynamics in the gauge $A^0=0$, whereas for $t$ being 
replaced by $\x$ and the special (singular) choice of gauge
\ben
A^0(\g) := A^\x(\g) = 
\frac{2A^\g(1)-\g(1+\g)A^\g(\g)}
{(\x-\xb)(1-\g)}
\een
one exactly recovers the equations of motion (\ref{A1}).

Finally the surviving first-class constraints (\ref{con}) that are due
to former flatness on the sphere gain a definite physical
meaning in the principal model of dimensionally reduced
gravity. Arising there equivalently as regularity conditions in
$\g\!\sim\!\infty$ \Ref{cona}, they are directly related to the
asymptotical flatness of the corresponding solution $g$ of Einstein's
equations \Ref{ee}. As first-class constraints in different pictures
\cite{BreMai87}, they generate respectively the Matzner-Misner
or the Ehlers symmetry transformations of the model.

Their actual role as physical gauge transformation related to the
local Lorentz transformations becomes manifest in the proper treatment
of the coset model below, see subsection \ref{fs}.

\subsection{The algebra of observables}\la{algob}

A consistent treatment of the theory and in particular the ability to
extract classical and quantum predictions from the theoretical
framework requires the identification of a complete set of
observables. In the model as presented so far, observables can be
defined in the sense of Dirac as objects that have vanishing Poisson
bracket with all the constraints including the Hamiltonian constraints
(\ref{C}), which even play the most important role here. In two-time
formalism this condition shows the observables to have no total
dependence on $\x$ and $\xb$. This is a general feature of a covariant
theory, where time dynamics is nothing but unfolding of a gauge
transformation, and observables are the gauge invariant objects.

Regarding the connection $A(\g)$ as fundamental variables of the
theory, the natural objects to build observables from are the
monodromies of the linear system (\ref{A}). They may be equivalently
characterized as
\be
\P(\g)\mapsto\P(\g)M_{l} \qquad 
\mbox{for $\g$ running along the closed path $l$}
\ee
or
\ben
M_{l} = P \exp\left(\oint_{l}A(\g){\rm d}\g\right).
\een

These objects naturally have no total ($\x,\xb$)-dependence; in the
isomonodromic sector we treat, the $w$-dependence is also absent.

For simple poles let us denote by $M_i \equiv M_{l_i}$ the monodromies
corresponding to the closed paths $l_i$ which respectively encircle
the singularities $\g_i$ and touch in one common basepoint. From the
local behavior of $\P(\g)$ around $\g=\g_i$:
\ben
\P(\g)= G_i\Big(I+\cO(\g-\g_i)\Big)(\g-\g_i)^{T_i}C_i
\een
one also extracts the relations 
\be\label{AM}
A_i=G_iT_iG_i^{-1}, \quad
M_i=C_i^{-1}{\rm e}^{2\pi iT_i}C_i.
\ee

The remaining constraint of the theory which should have vanishing
Poisson bracket with the observables is the generator of the constant
gauge transformations (\ref{con}), under which the monodromies
transform by a common constant conjugation. This justifies
\begin{Definition}\la{Obs}
In the case, where the connection $A(\g)$ exhibits only simple poles
at fixed singularities $w_j$ and with fixed eigenvalues of $A_j$, we
call the set of Wilson loops
\be\label{obs}
\{\tr \prod_k M_{i_k}|k,(i_1,\dots,i_k) \} 
\ee
the {\bf set of observables}.
\end{Definition}

\begin{Note}\la{ms}\rm
For these connections $A(\g)$, the corresponding monodromies together
with the position of the singularities and the eigenvalues of $A_j$
generically already carry the complete information.  (It is necessary
to add the set of eigenvalues of $A_j$ --- i.e.~the matrices $T_j$ or
the Casimir operators of the algebra respectively --- to the set of
monodromies, since from the monodromies only the exponentials of these
eigenvalues can be extracted.)
In the presence of higher order poles in the connection, additional
scattering data --- so-called Stokes multipliers --- are required to
uniquely specify the connection \cite{JiMiUe81}.

The generic case, in which the whole information is contained in the
above data, is precisely defined by the fact that no eigenvalues of
the monodromy matrices coincide \cite{JiMiMoSa80,JiMiUe81}. In
particular, this excludes the case of multisolitons, where the
monodromies equal $\pm I$.
\end{Note}

The algebraic structure of the observables (\ref{obs}) is 
inherited from the Poisson structure on the corresponding connection
$A(\g)$.

Before we explicitly describe this structure, let us briefly comment
on the relation to Chern-Simons theory, where quite similarly the
Poisson bracket (\ref{ABbracket}) provides a Poisson structure on
gauge invariant objects.

\subsubsection{Observables in Chern-Simons theory}
In Chern-Simons theory on the punctured sphere, the set of observables
is also built from the monodromy matrices. Note that since in the
usual gauge $A^0=0$ the Hamiltonian constraint is absent, observables
are identified as gauge invariant objects, where this is invariance
under local ($\g$-dependent) gauge transformations. Fixing this gauge
freedom by holomorphic gauge as described above, the Dirac bracket
(\ref{hbracket}) is now a structure on the reduced phase space of
holomorphic connections $A(z)$ modulo the action of {\em constant}
gauge transformations.

It has been explained in \cite{AlGrSc95a}, that the canonical bracket
(\ref{ABbracket}) does not define a unique structure on monodromy
matrices due to arising ambiguities from the singularities of this
bracket (see also \ci{Seme93}). However, on gauge invariant objects,
built from traces of arbitrary products of monodromy matrices, these
ambiguities vanish \cite{FocRos92, Alek93}. Hence the strategy there
is to postulate some structure on the monodromy matrices which reduces
to the proper one \ci{Gold86} on gauge invariant objects.

The holomorphic Dirac bracket (\ref{hbracket}) allows the calculation
also for the monodromies themselves, as we shall show in the
following.  To relate this result to \cite{FocRos92,AlGrSc95a}, note
that in general the original Poisson bracket and reduced Dirac bracket
coincide on quantities of first class in Dirac terminology, i.e.~here
on gauge invariant objects.  In this sense the holomorphic reduction
finally leads to the same result on the space of observables.

\subsubsection{Poisson structure of monodromy matrices}

The holomorphic Poisson structure (\ref{hbracket}) defines a Poisson
structure on the monodromy matrices $M_j$. The result is summarized in
the following

\begin{Theorem}\label{mono}
Let $A(\g)$ be a connection on the punctured plane $\g / \{\g_1, \dots, 
\g_N\}$, equipped with the Poisson structure:

\be
\left\{A(\g)^0,A(\m)^{\bar{0}}\right\} = 
\frac1{\g-\m}\,\left[\,\O, A(\g)^0 + A(\m)^{\bar{0}}\right]
\label{PoissonA}
\ee
Let further $\Psi$ be defined as solution of the linear system
\be
\partial_\g\Psi(\g) = A(\g)\Psi(\g),
\ee
normalized at a fixed basepoint $s_0$
\be
\Psi(s_0) = I\label{norm}
\ee
and denote by $M_1, \dots, M_N$ the monodromy matrices of $\Psi$
corresponding to a set of paths with endpoint $s_0$, which encircle
$\g_1, \dots, \g_N$, respectively. Ensure holomorphy of $\Psi$ at
$\infty$ by the first-class constraint
\be\label{con2}
A_\i=\lim_{\g\rightarrow\infty} \g A(\g) = 0 .
\ee 
Then, in the limit $s_0\!\rightarrow\! \infty$, the Poisson structure 
of the monodromy matrices is given by:
\ba
\left\{M_i^0, M_i^{\bar 0}\,\right\} &=& 
i\pi\,\Big( M_i^{\bar 0}\,\O \,M_i^0 - \,M_i^0\,\O \,M_i^{\bar 0}
\Big)\label{monoMiMi}\\
\left\{M_i^0,M_j^{\bar 0}\,\right\} &=& 
i\pi \,\Big( M_i^0\,\O\,M_j^{\bar 0} + M_j^{\bar 0}\,\O\,M_i^0  -
\O\,M_i^0M_j^{\bar 0} - M_i^0M_j^{\bar 0}\,\O\Big) 
\quad\enspace {\it for }\enspace i<j, \label{monoMiMj} 
\ea
where the paths defining the monodromy matrices $M_i$ are ordered with 
increasing $i$ with respect to the distinguished path 
$[s_0\!\rightarrow\! \infty]$.
\end{Theorem}

At this point several comments on the result of this theorem are in order, 
whereas the proof is postponed to appendix \ref{pmono}.

\begin{Note} \rm
The first-class constraint (\ref{con2}) generates constant gauge
transformations of the connection $A$ in the Poisson structure
(\ref{PoissonA}). For the connections of the type (\ref{Ahp}) this
reduces to the constraint (\ref{con}). In terms of the monodromy
matrices, holomorphy of $\Psi$ at $\infty$ is reflected by
\be
M_\i\equiv \prod M_i = I, \label{conM}
\ee 
which in turn is a first-class constraint and generates the action of
constant gauge transformations on the monodromy matrices in the
structure (\ref{monoMiMi}) and (\ref{monoMiMj}). The ordering of this
product is fixed to coincide with the ordering that defines
(\ref{monoMiMj}).

The gauge transformation behavior of the fields explicitly reads
\ba
\left\{A^0_\i \,,\,A_j^{\bar 0}\,\right\} &=& 
\left[\,\O\,,\,A_j^{\bar 0}\right] \la{gtbA}\\
\left\{ M^0_\i\,,\,M_j^{\bar 0}\,\right\} &=&
i\pi  
\Big( M_\i^0\,\O\,M_j^{\bar 0} - M_j^{\bar 0}\,\O\,M_\i^0  -
\O\,M_\i^0M_j^{\bar 0} + M_\i^0 M_j^{\bar 0}\,\O\Big) 
\la{gtbM}
\ea
This transformation law is further inherited by arbitrary products
$M=\prod_k M_{j_k}$ of monodromies, where on the constraint surface
$M_\i=I~$ it takes the form
\be\la{gtbM2}
\left\{ M_\i^0, M^{\bar 0}\,\right\}= -2\pi i\left[\,\O\,,\,M^{\bar
0}\right],
\ee
resembling \Ref{gtbA}.

The generators of gauge transformations build the algebra
\be
\left\{ A_\i^0, A_\i^{\bar 0}\,\right\}= \left[ \,\O, A_\i^{\bar 0}\right]
\la{Ag}
\ee
or
\be
\left\{M_\i^0,\; M_\i^{\bar 0}\,\right\} = 
i\pi\,\Big( M_\i^{\bar 0}\,\O \,M_\i^0 - \,M_\i^0\,\O \,M_\i^{\bar 0}
\Big) \la{Mg}
\ee
in terms of $A_\i$ and $M_\i$ respectively. In fact, the algebras
\Ref{Ag} and \Ref{Mg} turn out to be isomorphic: the quadratic bracket
\Ref{Mg} linearizes if the Casimirs are split out.

As mentioned, we will further be interested in gauge invariant
objects, which are now identified by their vanishing Poisson bracket
with (\ref{conM}) and which are therefore invariant under a global
common conjugation of all monodromies. Note, that this includes
invariance under gauge transformations with gauge parameters
(conjugation matrices) that have nonvanishing Poisson bracket with the
monodromies themselves. In accordance with Definition \ref{Obs}, the
structure \Ref{monoMiMi}, \Ref{monoMiMj} implies
\be
\left\{ M_\i, \tr M\right\} =0
\ee
for an arbitrary product of monodromies $M$.
\end{Note}

\begin{Note}\label{eyelash}\rm
The evident asymmetry of (\ref{monoMiMj}) with respect to the
interchange of $i$ and $j$ is due to the fact, that the monodromy
matrices are defined by the homotopy class of the path, which connects
the encircling path with the basepoint in the punctured plane. This
gives rise to a cyclic ordering of the monodromies.

The distinguished path $[s_0\!\rightarrow\! \infty]$ breaks and
thereby fixes this ordering, as is explicitly illustrated in figure
\ref{mono2} in appendix \ref{pmono} below. It is remnant of the
so-called eyelash that enters the definition of the analogous Poisson
structure in the combinatorial approach
\cite{FocRos92,Alek93,AlGrSc95a}, being attached to every vertex and
representing some freedom in this definition. However, the choice of
another path $[s_0\!\rightarrow\! \infty]$ simply corresponds to a
global conjugation by some product of monodromy matrices: a shift of
this eyelash by $j$ steps corresponds to the transformation
\ben
M_k\rightarrow (M_1\dots M_j)^{-1} M_k  (M_1\dots M_j)
\een
Therefore the restricted Poisson structure on gauge invariant objects
is independent of this path.
\end{Note}

\begin{Note}\rm\la{monor}
A seeming obstacle of the structure (\ref{monoMiMi}), (\ref{monoMiMj})
is the violation of Jacobi identities. Actually, this results from
heavily exploiting the constraint (\ref{con2}) in the calculation of
the Poisson brackets. As therefore these brackets are valid only on
the first-class constraint surface (\ref{conM}), Jacobi identities can
not be expected to hold in general.

However, the same reasoning shows \ci{Sche97}, that the structure
(\ref{monoMiMi}), (\ref{monoMiMj}) restricts to a Poisson structure
fulfilling Jacobi identities on the space of gauge invariant
objects. On this space, the structure reduces to the original Goldman
bracket \ci{Gold86} and coincides with the restrictions of previously
found and studied structures on the monodromy matrices
\cite{FocRos92}:
\ba
\left\{M_i^0 \, ,\, M_i^{\bar 0}\,\right\} &=& 
M_i^{\bar 0}r_+M_i^0 + M_i^0r_-M_i^{\bar 0} -
r_-M_i^0M_i^{\bar 0}-M_i^0M_i^{\bar 0}r_+ 
 \label{monorMM}\\
\left\{M_i^0 \, , \,  M_j^{\bar 0}\,\right\} &=& 
M_i^0r_+ M_j^{\bar 0} + M_j^{\bar 0}r_+ M_i^0 -
r_+M_i^0M_j^{\bar 0} - M_i^0M_j^{\bar 0}r_+   
\qquad {\rm for }\enspace i<j, \nn 
\ea
where $r_+$ and $r_-\!:=\!-\Pi r_+ \Pi$ are arbitrary solutions of
the classical Yang-Baxter equation
\be\label{cYB}
[r^{12},r^{23}] + [r^{12},r^{13}] + [r^{13},r^{23}] = 0.
\ee
and the symmetric part of $r_+$ is required to be $i\pi\O$. Setting
$r_+ \equiv i\pi\O$, (\ref{monorMM}) reduces to (\ref{monoMiMi}),
(\ref{monoMiMj}) such that our structure is in some sense the
skeleton, which may be dressed with additional freedom that vanishes
on gauge invariant objects. On the space of monodromy matrices
themselves, introduction of $r$-matrices may be considered as some
regularization to restore associativity, whereas the fact that $\O$
itself does not satisfy the classical Yang-Baxter equation is
equivalent to (\ref{monoMiMi}), (\ref{monoMiMj}) not obeying Jacobi
identities.

In the Poisson structure \Ref{monorMM}, the generator of gauge
transformations $M_\i\equiv\prod_i M_i$ has the following Poisson
brackets with any monodromy $M_k$:
\be
\left\{M_\i^0,M_k^{\bar 0}\,\right\}=M^{\bar 0}_k r_+ M_\i^0 - 
M^{\bar 0}_k M_\i^0 r_- -
r_+  M_\i^0 M^{\bar 0}_k + M_\i^0 r_- M^{\bar 0}_k  
\ee
which entails the same Poisson bracket of $M_\i$ with an arbitrary
product of monodromies $M\equiv \prod_k M_{j_k}$. On the constraint
surface $M_\i = I$, taking into account $r_+\!-\!r_- =2i\pi\O$, this
again implies \Ref{gtbM2}, such that $M_\i$ again generates the
constant gauge transformations.
\end{Note}

\begin{Note}\rm\la{tr}
The subset of observables 
\be
\{\tr [(M_i)^m]| i,m\} \cup \{w_i |i\}
\ee
commutes with the whole set of observables.

For the positions of the singularities this follows just 
trivially from the Poisson structure (\ref{PB1}), whereas the
eigenvalues of the monodromy matrices are related to the eigenvalues
of the corresponding residues $A_i$ (\ref{AM}), which in turn provide
the Casimir operators of the mutually commuting algebras
(\ref{PBAsp}). This subset of commuting variables thus parametrizes
the symplectic leaves of \Ref{monoMiMi},\Ref{monoMiMj}.
\end{Note}

\begin{Note}\rm
For our treatment of the coset model below, the following additional
structure will be of importance. There is an involution $\tilde{\n}$
on the set of observables, defined by the cyclic shift $M_i\mapsto
M_{i\pm n}$, where $N=2n$ is the total number of monodromies. The
crucial observation is now, that this involution is an automorphism of
the Poisson structure on the algebra of observables:
\be
\{\tilde{\n}(X_1),\tilde{\n}(X_2)\} = \tilde{\n} ( \{X_1,X_2\} ) , 
\ee
for $X_1, X_2$ being traces of arbitrary products of monodromy
matrices. This is a corollary of Note \ref{eyelash}, as it 
follows from the invariance of the Poisson structure on gauge
invariant objects with respect to a shift of the eyelash that defines
the ordering of monodromy matrices.

Like every involution, $\tilde{\n}$ defines a grading of the
algebra into its eigenspaces of eigenvalue $\pm 1$. In particular, the
even part forms a closed subalgebra.
\la{etat}
\end{Note}

\section{Quantization of the principal model}
\subsection{Quantization in terms of the connection}

The quantization of the model looks especially natural in the
isomonodromic sector with only simple poles. This has been performed
in \cite{KorNic95b,KorNic96}, as we shall briefly summarize.  In this
case straightforward quantization of the linear Poisson brackets
\Ref{PBAsp} leads to the following commutation relations:
\ba
[A_i^a,A_j^b] &=& i\hbar\d_{ij}f^{abc}A_j, \la{CR}\\
{}[ \x, (\log h)_\x  ] &=& [\xb, (\log h)_{\xb} ] ~=~ -i\hbar,
\label{CR1}\\ 
{}[ \xb, (\log h)_\x  ] &=& [\x, (\log h)_\xb ] ~=~ 0\nn
\ea
According to \Ref{CR1}, representing $\x$ and $\xb$ by multiplication 
operators, one can choose
\be\la{hq} 
(\log h)_{\x} =  i\hbar \f{\p}{\p \x}\qquad
(\log h)_{\xb} =  i\hbar \f{\p}{\p \xb}
\ee

{}From \Ref{CR}, the residues $A_j$ can be represented according to
\be\la{Aj}
A_j^a = i\hbar t^a_j, 
\ee
which acts on a representation $V_j$ of the algebra
${\mathfrak g}_\C$.

Thus the quantum state $\psi(\x,\xb)$ in a sector with given
singularities should depend on $(\x,\xb)$ and live in the
tensor-product
\ben
V^{(N)}:=V_1\otimes\dots\otimes V_N
\een
of $N$ representation spaces. Denote the dimension of $V_j$ by
$d_j$, such that $d\!:=\!{\rm dim}V^{(N)}\!=\!\prod d_j$.

\subsubsection{Wheeler-DeWitt equations and Knizhnik-Zamolodchikov
system} 

The whole ``dynamics'' of the theory is now encoded in the constraints
\Ref{C}, which accordingly play the role of the Wheeler-DeWitt
equations here:
\be\la{WDW1}
\Cx\psi= \Cxb\psi = 0
\ee
which can be written out in explicit form using \Ref{C}, \Ref{H},
\Ref{hq} and \Ref{Aj}:
\ba
\f{\p\psi}{\p\x}&=& 
\frac{i\hbar}{\x-\xb} 
\sum_{k \neq j} \f{\Omega_{jk}}{(1-\g_j)(1-\g_k)}\;\psi\la{WDW2}\\
\f{\p\psi}{\p\xb}&=& 
\frac{i\hbar}{\xb-\x} 
\sum_{k\neq j} \f{\Omega_{jk}}{(1+\g_j)(1+\g_k)}\;\psi\nn
\ea
where $\Omega_{jk} := t_j^a\otimes t_k^a$ is the symmetric 2-tensor of
${\mathfrak g}$, acting nontrivially only on $V_j$ and $V_k$.

The other constraint that restricts the physical states arrives from
\Ref{con}; its meaning was sketched in subsection \ref{embedd}. In the
quantized sector it is reflected by:
\be\la{conq}
\left(\sum_j t_j^a\right) \psi(\x,\xb) = 0
\ee

The general solution of the system \Ref{WDW2} is not known. 
However, these equations turn out to be intimately related to the
Knizhnik-Zamolodchikov system \cite{KniZam84}:
\be
\f{\p \varphi_{\scriptscriptstyle KZ}}{\p\g_j}=i\hbar\sum_{k\neq
j}\f{\Omega_{jk}} {\g_j-\g_k} \varphi_{\scriptscriptstyle KZ}
\la{KZ}\ee
with an $V^{(N)}$-valued function $\varphi_{\scriptscriptstyle
KZ}(\g_j)$:

\begin{Theorem}\la{KZWDW}
If $\varphi_{\scriptscriptstyle KZ}$ is a solution of \Ref{KZ} obeying
the constraint \Ref{conq}, and the $\g_j$ depend on $(\x,\xb)$
according to \Ref{gamma}, then
\be\la{link}
\psi = 
\prod_{j=1}^{N} \bigg( \f{\partial \g_j}{\partial w_j}
      \bigg)^{\frac12 i\hbar\O_{jj}} 
\varphi_{\scriptscriptstyle KZ} 
\ee
solves the constraint (Wheeler-DeWitt) equations \Ref{WDW2}. 
\end{Theorem}
The Casimir operator $\O_{jj}$ defined above is assumed to act
diagonal on the states, for ${\mathfrak g}$=$\mathfrak{sl}(2)$ for
example, this is simply $\O_{jj}=\f12 s_j (s_j-2)$,
classifying the representation.

Theorem \ref{KZWDW} and the proof were obtained in
\cite{KorNic95b}.\qed

Thus, the task of solving \Ref{WDW2} reduces to the solution of
\Ref{KZ}.
\begin{Note}\la{gjdyn}\rm
The $\g_j$ dependence of the quantum states, introduced in Theorem
\ref{KZWDW}, can be understood as just a formal dependence, which
covers the ($\x,\xb$)-dependence of these states. However, one may also
split up this dynamics into several commuting flows generated by the
corresponding operators from \Ref{KZ}. In this sense then, the full
set of solutions of \Ref{KZ} may be interpreted as a
``$\g_j$-evolution operator'', describing this dynamics. In some sense
\ci{KorNic96} this quantum operator resembles the classical
$\t$-function introduced in \cite{JiMiMoSa80}.
\end{Note}

\begin{Note}\rm\la{qhipo}
We have described, how the solution of the Wheeler-De Witt equations
is related to the solution of the KZ system \Ref{KZ} in the sector of
the theory, where the connection has only simple poles. It is
therefore natural to suppose, that the quantization of the higher pole
sectors that were classically presented in subsection \ref{hp} is
achievable in a similar way and will moreover reveal a link to the
higher order KZ systems, which were introduced in \cite{Resh92} in the
quantization of isomonodromic deformations with exactly the Poisson
structure \Ref{PBAh} on the residues.
\end{Note}

\begin{Note}\rm\la{ising}
For definiteness it is convenient to assume pure imaginary
singularities $\g_j\!\in\!i\R$ (i.e.~$w_j\!\in\!\R$). Then classically
$A_j\in {\mathfrak g}$ and quantized they carry representations of
${\mathfrak g}$ itself, not of ${\mathfrak g}_\C$.
\end{Note} 

\subsection{Quantum algebra of monodromy matrices}
\subsubsection{Quantum monodromies}
Having quantized the connection $A(\g)$ as described in the previous
section, it is a priori not clear how to identify quantum
operators corresponding to the classical monodromy matrices in this
picture. As they are classically highly nonlinear functions of the
$A_j$, arbitrarily complicated normal-ordering ambiguities may arise
in the quantum case.

The first problem is the definition of the quantum analogue of the
classical $\Psi$--function. Its $d_0\!\times\!d_0$ matrix entries are
now operators on the $d$-dimensional representation space
$V^{(N)}$. We choose here a simple convention, replacing the classical
linear system
\be\la{lsq}
\p_\g \Psi(\g) = A(\g)\Psi(\g)
\ee
by formally the same one, where all the arising matrix entries are
operators now, i.e.~\Ref{lsq} remains valid for higher dimensional
matrices $A$ and $\Psi$. We have thereby fixed the operator ordering
on the right hand side in what seems to be the most natural way. In
the same way, we define the quantum monodromy matrices:
\begin{Definition}
The {\bf quantum monodromy matrix} $M_j$ is defined to be the
right-hand-side monodromy matrix of the (higher dimensional) {\bf
quantum linear system} \Ref{lsq}:
\be\la{Mq}
\Psi(\g)\mapsto\Psi(\g)M_j\qquad\mbox{for $\g$ encircling $\g_j$},
\ee
where the quantum $\Psi$-function is normalized as
\be\la{norm2}
\Psi(\g) = \left(I+\cO\left(\frac1{\g}\right)\right)\g^{-A_\i}\qquad
\mbox{around $\g \sim \infty$}.
\ee
\end{Definition}
\begin{Note}\rm
The normalization condition \Ref{norm2} generalizes the one we chose
in the classical case \Ref{norm} where the basepoint $s_0$ was sent to
infinity. This generalization is necessary, because the constraint
\Ref{con2} is not fulfilled as an operator identity in the quantum
case, which means, that the quantum $\Psi$-function as an operator is
definitely singular at $\g=\infty$ with the behavior \Ref{norm2}. Only
its action on physical states, which are by definition annihilated by
the constraint \Ref{con} may be put equal to the identity for
$\g=\infty$.
\end{Note}
For further proceeding we now make use of an interesting observation
of \cite{Resh92}, relating the KZ-systems with $N$ and $N\!+\!1$
insertions by means of the quantum linear system \Ref{lsq}. We state
this as
\begin{Theorem}\la{ResT} 
Let $\varphi(\g_1,\dots,\g_N)$ be the evolution operator of the KZ-system
\ben
\p_j\varphi = i\hbar
\sum_{k\not=j}\frac{\O_{jk}}{\g_j-\g_k}\varphi
\een
and $\Phi(\g_0,\dots,\g_N)$ be the corresponding evolution operator of
the KZ-system with an additional insertion at $N\!=\!0$. Then
$\Psi(\g_0,\dots,\g_N) := (I\otimes\varphi^{-1})\Phi$ satisfies the
following system of equations:
\ba
\p_0 \Psi &=& i\hbar\sum_{j=1}^N 
\frac{t_0^a\otimes(\varphi t_j^a\varphi^{-1})}{\g_0-\g_j} \Psi\la{Res}\\
\p_j \Psi &=& -i\hbar\frac{t_0^a\otimes(\varphi t_j^a\varphi^{-1})}
{\g_0-\g_j} \Psi \nn
\ea
\end{Theorem}
The proof is obtained by a simple calculation.\qed

Consider the relations \Ref{Res}. Together with the remarks of Note
\ref{gjdyn}, it follows that this $\Psi$ just obeys the proper quantum
linear system \Ref{lsq} in a Heisenberg picture: the ($\x,
\xb$)-dependence of the operators $A_j$ is generated by conjugation
with the evolution-operator $\varphi$. For the definition of the quantum
$\Psi$-function it is Heisenberg picture which provides the most
natural framework, as only in this picture implicit and explicit ($\x,
\xb$)-dependence of operators are treated more or less on the same
footing. Thus one may identify:
\ben
A_j = i\hbar t_0^a\otimes(\varphi t_j^a\varphi^{-1})
\een
The operators $t^a_0$ play the role of the classical representation
$t^a$ acting on the auxiliary space $V_0$, which is already required
for the formulation of the classical linear system. In this sense, the
KZ-system with $N\!+\!1$ insertions combines the classical linear
system with the quantum equations of motion, that are described by the
KZ-system with $N$ insertions. The additional insertion $\g_0$ then
plays the role of $\g$.  We shall use this link to gain information
about the algebraic structure of the quantum monodromy matrices.

\subsubsection{Quantum group structure}
We now start from the representation of the quantum $\Psi$-function
due to Theorem \ref{ResT}:
\be\la{PsiPhi}
\Psi(\g,\g_1,\dots,\g_N) = 
\Big((I\otimes\varphi^{-1}(\g_1,\dots,\g_N)\Big)
\Phi(\g,\g_1,\dots,\g_N)
\ee
This shows in particular, that the quantum monodromy matrices of the
principal model defined in \Ref{Mq} equal the corresponding
monodromies of the KZ-system with $N\!+\!1$ insertions. To obtain
their algebraic structure, we employ a deep result of Drinfeld about
the relation between the monodromies of the KZ-connection and the
braid group representations induced by certain quasi-bialgebras
\cite{Drin89b,Drin89a}. Before we state these relations, we have to
briefly describe the induced braid group representations.

The KZ-system that is of interest here, is
\ben
\p_j \Phi = i\hbar\sum_{k\not=j} \frac{\O_{jk}}{\g_j-\g_k} \Phi,
\een
with $j=0,\dots,N$, which, as explained, in a formal sense combines
the classical and the quantum degrees of freedom, the function $\Phi$
living in $V^{(N+1)}:=V_0\otimes V^{(N)}$. This system naturally
induces a representation of monodromy matrices, which may canonically
be lifted to a braid group representation \cite{Kass95}. However, for
our purpose, it is sufficient to remain on the level of the monodromy
representation, which we denote by $\r_{KZ}$.

We further have to briefly mention two algebraic structures, which are
standard examples for braided quasi-bialgebras, where for details and
exact definitions we refer to \cite{Drin89a,Kass95}. Let us denote by
$\U_\hbar$ the so-called Drinfeld-Jimbo quantum enveloping algebra
associated with ${\mathfrak g}$ \cite{Drin85,Jimb85}. This is a
braided bialgebra, which includes the existence of a comultiplication
$\D$, a counit $\e$ and a universal $R$-matrix
$R_\U\in\U_\hbar\!\otimes\!\U_\hbar$, obeying several conditions of
which the most important here is the (quantum) Yang-Baxter equation:
\be\la{qYB}
R_\U^{12} R_\U^{13} R_\U^{23} = 
R_\U^{23} R_\U^{13} R_\U^{12}
\ee

The matrix $R_\U$ can in principle be explicitly given, but is of a
highly complicated form. It is Drinfeld's achievement to relate
this structure to a braided quasi-bialgebra $\A_\hbar$, where the
nontriviality of the $R$-matrix is essentially shifted into an
additional element $\phi_\A\in\A_\hbar\!\otimes\!\A_\hbar\!\otimes\!
\A_\hbar$, the
so-called associator, which weakens the coassociativity. The
$R$-matrix of $\A_\hbar$ is simply $R_\A\!=\!e^{-\pi\hbar\O}$, where
$\O:=t^a\!\otimes\!  t_a$ is the symmetric 2-tensor of
${\mathfrak g}$. This $R$-matrix satisfies a weaker form of
\Ref{qYB}, the quasi-Yang-Baxter equation:
\be\la{QYB}
R_\A^{12}\phi_\A^{312}R_\A^{13}(\phi_\A^{-1})^{132}
R_\A^{23}\phi_\A^{123} =
\phi_\A^{321}R_\A^{23}(\phi_\A^{-1})^{231}R_\A^{13}
\phi_\A^{213}R_\A^{12}
\ee

The algebras $\U_\hbar$ and $\A_\hbar$ are isomorphic as braided
quasi-bialgebras \ci{Drin89a}.\bigskip

There is a standard way, in which braided quasi-bialgebras induce
representations of the braid group. Each simple braid $\s_i$ is
represented as
\be\la{monrep}
\r(\s_i):= \phi_i^{-1}\Pi^{i,i+1}R^{i,i+1}\phi_i,
\ee
where $\Pi$ is the permutation operator and $\phi_i$ is defined as
$\phi_i:=\D^{(i+1)}(\phi)\otimes I^{\otimes(N-i-2})$ with
$\D^{(1)}\!:=\!1$, $\D^{(2)}\!:=\!{\rm Id}$ and
$\D^{(i+1)}:=(\D\otimes{\rm Id}^{\otimes i})\D^{(i)}$. We will denote
the restrictions of these representations of the algebras $\U_\hbar$ and
$\A_\hbar$ on the monodromies, which are built from products of simple
braids, by $\r_\U$ and $\r_\A$ respectively.

Now we have collected all the ingredients to state the result of 
Drinfeld as:

\begin{Theorem}\la{DrinT}
The monodromy representation of the KZ-system equals the described
monodromy representation of the braided quasi-bialgebra $\A_\hbar$,
which in turn is equivalent to the monodromy representation of the
braided bialgebra $\U_h$. This means, that there is an automorphism
$u$ on $V^{(N+1)}$, such that
\be
\r_{\rm KZ} = \r_{\A} = u\r_{\U}u^{-1}
\ee
\end{Theorem}
For the proof we refer to the original literature \cite{Drin89a} or to
the textbook of Kassel \cite{Kass95}.

We should stress that in this construction the deformation parameter
of quantum group structure coincides with the true Planck constant
$\hbar$.
\qed

\subsubsection{Quantum algebra and classical limit}
It was our aim to describe the algebraic structure of the quantum
monodromy matrices defined in \Ref{Mq}. By Theorem \ref{ResT} these
monodromy matrices have been identified among the monodromies of the
KZ-system with $N\!+\!1$ insertions as the monodromies of the
additional point $\g_0$ encircling the other insertions. Exploiting
the consequences of Theorem \ref{DrinT} now, the quantum algebra of
the monodromy matrices $M_1,\dots, M_N$ is given by:

\begin{Theorem}\la{MQaT}
The matrices $M_j$ from (\ref{Mq}) satisfy
\ba\la{MQa}
R_-M_i^0R_-^{-1}M_i^{\bar 0} &=& M_i^{\bar 0}R_+M_i^0R_+^{-1}\\
R_+M_i^0R_+^{-1}M_j^{\bar 0} &=& M_j^{\bar 0}R_+M_i^0R_+^{-1},
\qquad \mbox{for $i<j$},\nn
\ea
where these relations are understood in a fixed representation of the
$d_0\times d_0$ matrix entries of the monodromy matrices on the
tensor-product $V^{(N)}=\bigotimes_j V_j$.
The $R$-matrices $R_\pm$ are given by
\be
R_- := u_{\bar 0}R_\U^{-1} u^{-1}_0,
\qquad R_+ := \Pi R_-^{-1} \Pi
\ee
where $R_\U$ is the universal $R$-matrix of $\U_\hbar$ mentioned
above, $u_0$ is some automorphism on $V_0\otimes V^{(N)}$ and $u_{\bar
0}$ is the corresponding one on $V_{\bar 0}\otimes V^{(N)}$. The
classical limit of these $R$-matrices is given by:
\be\la{cllim}
R_\pm = I\!\otimes\!I \,\pm\, (i\hbar)(i\pi\O) \,+\, \cO_\pm(\hbar^2)
\ee
\end{Theorem}
\begin{Note}\rm
The relations \Ref{MQa} are to be understood as follows. The notation
requires two copies $0$ and ${\bar 0}$ of the classical auxiliary
space $V_0$. While the standard $R$-matrices $R_\U$ and $R_\A$ live on
these classical spaces only, $R_-$ and $R_+$ also act nontrivially on
the quantum representation space $V^{(N)}$, due to conjugation with
the automorphisms $u_0, u_{\bar 0}$.
\end{Note}

\paragraph{Proof of Theorem \ref{MQaT}:}
Consider the monodromy representation \Ref{monrep} corresponding to
the coassociative bialgebra $\U$. The monodromy $M_j$ for
$\g\!=\!\g_0$ encircling $\g_j$ is thereby represented as:
\be
\r_\U(M_j) = 
(R_\U^{-1})^{01}(R_\U^{-1})^{02}\dots R_\U^{j0}
R_\U^{0j}R_\U^{0,j-1}\dots R_\U^{01}, 
\ee
such that it is just a matter of sufficiently often exploiting the
Yang-Baxter equation \Ref{qYB} to explicitly show, that the relations
\Ref{MQa} hold for $\r_\U(M_j)$ with $R_-:=R_\U^{-1}, R_+:=\Pi
R_-^{-1} \Pi$. Theorem \ref{DrinT} further implies the conjugation of
the $R$-matrices with the automorphism $u$ in order to extend the
result to the representation $\r_{\rm KZ}$, in which the monodromies
from \Ref{Mq} were recovered.

To further prove the asymptotic behavior \Ref{cllim}, it is not enough
to know the classical limit of $R_\U$ --- which is a classical
$r$-matrix simply ---, since the semiclassical expansion of the
automorphisms $u_0, u_{\bar 0}$ must be taken into account. For this
reason, we additionally have to use the other part of Theorem
\ref{DrinT}, which relates the representations $\r_{\rm KZ}$ and
$\r_\A$. The relations \Ref{MQa} for the $\r_\A(M_j)$ hold with
$R_-:=R_\A^{-1}, R_+:=\Pi R_-^{-1}\Pi$ in a generalized form, modified
by certain conjugations with the nontrivial associator $\phi_\A$. The
semiclassical expansion of the associator is given by \cite{Kass95}:
\be\la{limas}
\phi_\A = I\!\otimes\!I\!\otimes\!I \,+\, \cO(\hbar^2)
\ee
which implies, that the term of order $\hbar$ in the semiclassical
expansion of \Ref{MQa} is determined by the corresponding one in
$R_\A=e^{-\pi\hbar\O}$, which yields \Ref{cllim}. 

The last point to be ensured is, that the normalization of the quantum
monodromies \Ref{norm2} around $\g\sim\infty$ coincides with the
normalization chosen in the definition of the KZ-monodromies
\cite{Drin89b} in certain asymptotic regions of the space of
$(\g,\g_1\,\dots,\g_N)$, up to the order $\hbar$. The proof of this
fact goes along the same line as the proof of \Ref{limas}.\qed

We have now established the quantum algebra of the quantum monodromy
matrices by identifying the corresponding operators inside the picture
of the quantized holomorphic connection $A(\g)$.  The classical limit
of this algebra equals exactly the classical algebra of monodromy
matrices \Ref{monoMiMi}, \Ref{monoMiMj}. Hence, we have shown the
``commutativity'' of the (classical and quantum) links between the
connection and the monodromies with the corresponding quantization
procedures. Let us sketch this in the following diagram:
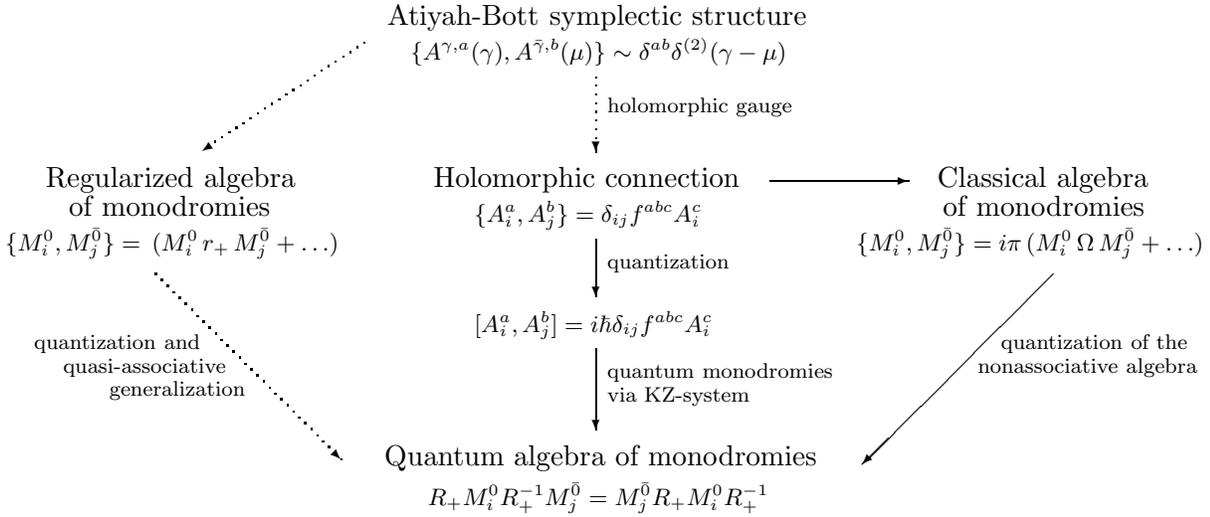
\begin{center}
 \setlength{\unitlength}{0.2em}
 \begin{picture}(160,95)
 \put(47,55){Holomorphic connection}
 \put(54,49){\footnotesize $\{A^a_i,A^b_j\} = 
   \d_{ij}f^{abc}A^c_i$}
 \put(75,46){\vector(0,-1){10}}
 \put(77,41){\scriptsize quantization}
 \put(54,30){\footnotesize $[A^a_i,A^b_j] = 
     i\hbar\d_{ij}f^{abc}A^c_i$}
 \put(75,27){\vector(0,-1){14}}
 \put(77,22){\scriptsize quantum monodromies}
 \put(77,18){\scriptsize via KZ-system}
 \put(38.5,7){Quantum algebra of monodromies}
 \put(46,0){\footnotesize $R_+M_i^0R_+^{-1}M_j^{\bar 0} = 
            M_j^{\bar 0}R_+M_i^0R_+^{-1}$}
 \put(105,56){\vector(1,0){24}}
 \put(134,55){Classical algebra}
 \put(135,50){of monodromies}
 \put(120,44){\footnotesize $\{M_i^0,M_j^{\bar 0}\} = 
  i\pi \,( M_i^0\,\O\,M_j^{\bar 0} + \dots)$}
 \put(154,40){\vector(-1,-1){33}}
 \put(145,27){\scriptsize quantization of the}
 \put(141,23){\scriptsize nonassociative algebra}
 \put(75,62){\vector(0,-1){1}}
 \dottedline{1.5}(75,74)(75,62)
 \put(39,83){Atiyah-Bott symplectic structure}
 \put(43,77){\footnotesize $\{A^{\g,a}(\g),A^{\gb,b}(\m)\} \sim 
 \d^{ab}\d^{(2)}(\g-\m)$}
 \put(77,68){\scriptsize holomorphic gauge}
 \put(-20,55){Regularized algebra}
 \put(-16,50){of monodromies}
 \put(-27,44){\footnotesize $\{M_i^0,M_j^{\bar 0}\} = 
    \,( M_i^0\,r_+\,M_j^{\bar 0} + \dots)$}
 \put(8,62){\vector(-3,-2){1}}
 \dottedline{1.5}(8,62)(35,80)
 \dottedline{1.5}(-1,40)(31,8)
 \put(30,9){\vector(1,-1){1}}
 \put(-22,27){\scriptsize quantization and}
 \put(-17,23){\scriptsize quasi-associative}
 \put(-9,19){\scriptsize generalization}
  \end{picture}
\end{center}
 
\vspace{0.6cm}

\begin{Note}\rm\la{cw}
The dotted lines in this diagram depict the link to the usual way,
quantum monodromies have been treated. This was done by directly
quantizing their classical algebra, which is derived from the original
symplectic structure of the connection up to certain degrees of gauge
freedom: for later restriction on gauge invariant objects, this
algebra may be described with an arbitrary classical $r$-matrix, as
was sketched in Note \ref{monor}. A direct quantization of this
structure is provided by a structure of the form \Ref{MQa}, where the
quantum $R$-matrices live in the classical spaces only and admit the
classical expansion $R_\pm=I+i\hbar r_\pm + \cO_\pm(\hbar^2)$
\cite{Alek93,AlGrSc95a}.
\end{Note}
\begin{Note}\rm
In contrast to this quantum algebra which underlies \Ref{monorMM}, in
\Ref{MQa} the $R$-matrices --- due to the automorphisms $u_0, u_{\bar
0}$ --- also act nontrivially on the quantum representation
space. Their classical matrix entries may be considered as
operator-valued, meaning, that the quantum algebra can be treated
alternatively as nonassociative or as ``soft''. This is in some sense
the quantum reason for the fact, that the classical algebra
\Ref{monoMiMi}, \Ref{monoMiMj} fails to satisfy Jacobi identities.
However, note that \Ref{MQa} only describes the $R$-matrix in any
fixed representation of the monodromies; for a description of the
abstract algebra, compare the quasi-associative generalization in
\cite{AlGrSc95a,AlGrSc95b}, which provides the link between the
quantum structure described in the previous note and \Ref{MQa}. 
\end{Note}

\subsubsection{Quantum observables}

Let us discuss now the quantum observables, i.e.~operators commuting
with all the constraints. In analogy with the classical case it is
clear that all monodromies of the quantum linear system \Ref{Mq}
commute with the Hamiltonian constraints. Therefore, it remains to get
rid of the gauge freedom \Ref{conM}, i.e.~to identify functions of
monodromies commuting with quantum generators of the gauge
transformations. In the classical case the gauge transformations were
generated by matrix entries of the matrix $A_\i$ or, equivalently, of
the matrix $M_\i\!-\!I$.
The straightforward quantization of the classical algebra of gauge
transformations generated by $A_\i$ \Ref{Ag} is
\be
[A_\i^a, A_\i^b] = f^{ab}_c A_\i^c,
\la{Ainf}\ee
i.e.~coincides with ${\mathfrak g}$. In terms of $M_\i$, the algebra
of the same gauge transformations according to \Ref{MQa} reads

\be
R_- M_\i^0 R_-^{-1} M_\i^{\bar 0} = M_\i^{\bar 0} R_+ M_\i^0 R_+^{-1}
\la{Minf}\ee

The set of quantum observables is characterized as the set of
operator-valued functions $F$ of components of monodromies $M_j$
which commute with all components of $A_\i$:
\be
[F(\{M_j\}), A_\i^a]=0
\ee
Remind that in classical case observables were just traces of
arbitrary products of monodromies $M_j$. At the moment the quantum
analog of this representation is not clear. One should suppose that
there is a similar situation to the case we would have arrived at by
directly quantizing the algebra of monodromies, mentioned in Note
\ref{cw}.

In this case, which has been studied in the combinatorial quantization
of Chern Simons theory \ci{AlGrSc95a,AlGrSc95b}, the $R$-matrices live
in the classical spaces only and the transformation behavior of
arbitrary products of monodromies $M$ under gauge transformations
generated by $M_\i$ reads:
\ben
R_-M^0R_-^{-1}M_\i^{\bar 0} = M_\i^{\bar 0}R_+M^0R_+^{-1}
\een

Introducing the quantum trace $\tr_q M$ with characteristic
relations
\be
\tr_q^0 R^{0{\bar 0}} M^0  (R^{0{\bar 0}})^{-1}=\tr_q M^0
\ee
we see that the operators $\tr_q M$ commute with the components of
$M_\i$:
\be
[\tr_q M, M_\i^0] =0
\ee

Therefore, the quantum group generated by $M_\i$:
\be
R_- M_\i^0 R_-^{-1} M_\i^{\bar 0} =
M_\i^{\bar 0} R_+ M_\i^0 R_+^{-1}
\la{Minfq}\ee
in this approach plays the role of algebra of gauge transformations.

It appears a difference of this approach with the approach
which we mainly follow in this paper: instead of the Lie group $G$
generated by the algebra \Ref{Ainf}, the role of the gauge group is
played by its quantum deformation \Ref{Minfq}.  
A question therefore remains: what is the proper quantum gauge
group of a consistent quantum theory, the group $G$ itself or its
quantum deformation $G_q$?

\begin{Note}\rm
With the notation of the quantum trace at hand, the quantum analogue
of Note \ref{tr} can be formulated. From the abstract algebraic point
of view --- beyond the presented concrete representation of the
quantum monodromies ---, the quantum trace of powers of the $M_j$
build the center of the free algebra defined by \Ref{MQa} and may thus
be fixed according to the classical values.
\end{Note}

\section{Coset model}
In this final chapter we will explain, how to modify the previously
presented scheme in order to treat the coset models, which actually
arise from physical theories. The field $g$ is required to take values
in a certain representation system of the coset space $G/H$, where $H$
is the maximal compact subgroup of $G$.

This subgroup may be characterized by an involution $\n$ of $G$ as the
subgroup, which is invariant under $\n$. The involution can further be
lifted to the algebra ${\mathfrak g}$, e.g.~$\n(X)\!=\!-X^t$ for
$X\!\in\!{\mathfrak g}\!=\!\mathfrak{sl}(N)$. The algebra ${\mathfrak
g}$ is thereby split into its eigenspaces with eigenvalues $\pm 1$,
which are denoted by ${\mathfrak g} = {\mathfrak h}\oplus{\mathfrak
k}$, the subgroup $H$ underlying ${\mathfrak h}$.  In terms of the
involution, the field $g$ is restricted to satisfy:
\be\la{coset}
g\n(g)=I,
\ee
which defines the special choice of a representation system of the
coset space.

\subsection{Classical treatment}

Classically speaking, the Poisson structure for the $G/H$-valued model
may be obtained from the previously described Poisson structure for
the principal $G$-valued model by implementing additional constraints. 

These constraints were discussed in detail in \ci{KorNic96} and may be
equivalently formulated in terms of the function $\Psi$ or of the
connection $A$:
\ba
\n\left(\Psi\left(\frac1{\g}\right)\right)^{-1} g^{-1} 
\Psi(\g) &=& C_0~,
\la{conPsi}\\
A(\g)+\f{1}{\g^2}g\n\left(A\left(\f{1}{\g}\right)\right) g^{-1} 
&=& 0\la{conA}
\ea

The first line is a consequence of \Ref{coset} with $C_0=C_0(w)$ from
\Ref{C0} also satisfying $C_0\n(C_0)\!=\!I$ now. Studying the
monodromies of $\Psi$ shows, that in the isomonodromic sector, $C_0$
must be gauged to a constant matrix, using the freedom of right hand
side multiplication of the solution of \Ref{ls}. This can be seen from
equation \Ref{consM} below. Derivation of \Ref{conPsi} with respect to
$\g$ then yields \Ref{conA}.

An unpleasant feature of these constraints is, that they explicitly
contain the field $g$, which in this framework is not among the
fundamental variables. To avoid this difficulty, it is convenient to
slightly modify the Hamiltonian formalism of the principal
model. Namely, let us relax the normalization condition $\Psi
(\g\!=\!\i)=I$, which was imposed in \Ref{norma} before and consider
the function $\Psit$ related to $\Psi$ by a $G$-valued gauge
transformation $\cV$ instead:
\be
\Psit:= \cV(\x,\xb)\Psi
\la{psit}
\ee
Then it is $\Psit(\g\!=\!\i)=\cV$ and $gC_0 = \cV^{-1}\Psit(\g\!=\!0)$,
such that the coset constraint \Ref{coset} may be rewritten as:
\be\la{constrV}
g=\cV^{-1}\eta(\cV)
\ee
\medskip

The modified function $\Psit$ now satisfies the linear system
\be
\f{d\Psit}{d\x}=\left(-\f{1+\g}{1-\g} P_+ + Q_+\right)\Psit,\qquad
\f{d\Psit}{d\xb}=\left(-\f{1-\g}{1+\g} P_- + Q_-\right)\Psit,
\la{lst}\ee
with $(\x,\xb)$-dependent matrices $P_\pm\in{\mathfrak k}$ and
$Q_\pm\in{\mathfrak h}$ which can be reconstructed from $\cV$ on the
coset constraint surface \Ref{constrV}:
\ben
\cV_\x \cV^{-1} = P_+ + Q_+,\qquad
\cV_\xb \cV^{-1} = P_- + Q_-
\een

\begin{Note}\rm
In the coset model the M\"obius symmetry \Ref{Mosy} appears in
especially simple form \ci{BaiSas82}:
\ben
\cV\mapsto \Psit(\g)\,,\qquad
P_+\mapsto
\sqrt{\f{w-\xb}{w-\x}} P_+\,,\qquad
P_-\mapsto
\sqrt{\f{w-\x}{w-\xb}} P_-\,,\qquad
h\mapsto h
\een
\end{Note}
\medskip

In complete analogy to the principal model, we further introduce
\begin{Definition}
Define the connection $\At$ by:
\be\la{At}
\At(\g):=\p_{\g}\Psit(\g)\Psit^{-1}(\g).
\ee
\end{Definition}
\medskip

The constraint of regularity at infinity then reads:
\be\la{conat}
\At_\i:=\lim_{\g\rightarrow\i}\g\At(\g) = 0
\ee

The relations \Ref{cur} between the original fields and the connection
$\At$ now take the following form:
\be\la{cur2}
\left.\f{1}{\x-\xb} \At(\g,\x,\xb)\right|_{\g=1} 
= -P_+, \qquad
\left.\f{1}{\x-\xb} \At(\g,\x,\xb)\right|_{\g=-1}
= -P_-. 
\ee
Hence, the coset constraints \Ref{constrV} are equivalent to
\be\la{constrv}
\At(\pm 1)=-\eta\left(\At(\pm 1)\right)
\ee
which is implied by \Ref{conA}. Let us stress again, that the
originally equivalent coset constraints \Ref{coset}, \Ref{constrV} or
\Ref{constrv} are lifted to \Ref{conA} due to the special choice of
$C_0\!=\!const$ in the isomonodromic sector.

The constraints \Ref{conPsi} and \Ref{conA} take simpler forms in
terms of the new variables $\Psit$ and $\At$, since the field $g$ is
absorbed now:
\ba \n\left(\Psit\left(\frac1{\g}\right)\right)^{-1}\Psit(\g) &=& C_0
\la{conPsit}\\
\At(\g)+\frac1{\g^2}\n\left(\At\left(\frac1{\g}\right)\right) &=& 0
\la{conAt} 
\ea
The first of these equations is a sign of the invariance of the linear
system \Ref{lst} on the coset constraint surface under the extended
involution $\n^\i$, introduced in \cite{BreMai87}:
\be
\n^\i(\Psit(\g)):=\n\left(\Psit\left(\frac1{\g}\right)\right),
\la{eta}\ee
but is difficult to handle due to the unknown matrix $C_0$. The latter
form \Ref{conAt} of the constraint admits a complete treatment as
will be described below. Note, that the constraint of regularity at
infinity \Ref{conat} is already contained in \Ref{conAt} and is
thereby naturally embedded in the coset constraints.

The set of constraints \Ref{conAt} is complete and consistent in the
following sense:
\begin{Lemma}\la{tind}
The coset constraints \Ref{conAt} are invariant under ($\x,
\xb$)-translation on the constraint surface.
\end{Lemma}
\paragraph{Proof:}
The total $\x$-dependence of $\At$ can be extracted from \Ref{A1} to be
\ba
\f{\rm d}{{\rm d}\x} \At(\g) &=& \cV[A^{\x}(\g), A(\g)]\cV^{-1} 
+[\cV_\x \cV^{-1},\At(\g)] + \cV\f{\p A^{\x}(\g)}{\p\g}\cV^{-1} \non
&=& \left[\frac{-2P_+}{1-\g},\At(\g)\right] + 
\left[(P_++Q_+),\At(\g)\right] \non
&&{}-\frac{2P_+}{(1-\g)^2} 
+\frac{\g^2-2\g-1}{(\x-\xb)(1-\g)^2}\At(\g)
-\frac{\g(1+\g)}{(\x-\xb)(1-\g)}\p_\g\At(\g)\nn
\ea

Together with $\f{\rm d}{{\rm d}\x}\left(f\left(\frac1{\g}\right)\right) =
\left(\f{\rm d}{{\rm d}\x}f\right)\left(\frac1{\g}\right)$ for any function
$f(\g)$, which follows from the structure of $\g_\x$, a short
calculation reveals, that on the constraint surface \Ref{conAt} it is
\ben
\f{\rm d}{{\rm d}\x}
\left(\At(\g)+\frac1{\g^2}\n\left(\At\left(\frac1{\g}\right)\right)\right)
\approx -\g_\x\; \f{\rm d}{{\rm d}\g}
\left(\At(\g)+\frac1{\g^2}\n\left(\At\left(\frac1{\g}\right)\right)\right)
\approx 0
\een
\qed

In a Hamiltonian formulation these constraints therefore have weakly
vanishing Poisson bracket with the full Hamiltonian, which is required
for a consistent treatment. Let us now briefly present the Hamiltonian
formulation of the coset model in terms of the new variables.

\subsubsection{Poisson structure and Hamiltonian formulation}
The definition of the connection $\At$ already implies the relation
\be
\At(\g)=\cV A(\g)\cV^{-1}, 
\ee
such that from \Ref{A1} one extracts the equations of motion for these
new variables:
\ba
\f{\p \At}{\p\x}&=& \cV[A^{\x}, A]\cV^{-1} + \cV\f{\p
A^{\x}}{\p\g}\cV^{-1} +[\cV_\x \cV^{-1},\At],\la{At1}\\ \f{\p
\At}{\p\xb}&=& \cV[A^{\xb}, A]\cV^{-1} + \cV\f{\p
A^{\xb}}{\p\g}\cV^{-1} +[\cV_{\xb}\cV^{-1},\At].\nn
\ea
\medskip

In analogy with the principal model, this motivates
\begin{Definition}
Define on $\At(\g)$ the following Poisson structure:
\be
\left\{\At^a(\g),\At^b(\m)\right\}_{\cV} = 
-f^{abc}\frac{\At^c(\g)-\At^c(\m)}{\g-\m},
\la{PB1t}
\ee
and denote by {\bf implicit time-dependence} the $(\x,\xb)$-dynamics,
that is generated by
\ba
\la{Ht}
\Ht^{\x} &:=&\frac1{\x-\xb}\tr \At^2(1) - 
\tr [\At_\i(\p_\x \cV\cV^{-1})],\\
\Ht^{\xb} &:=&\frac1{\xb-\x}\tr \At^2(-1) - 
\tr [\At_\i(\p_\xb \cV\cV^{-1})],\nn
\ea
on the constraint surface \Ref{conat}. The remaining explicit
time-dependence is then defined to be generated in analogy to
\Ref{PB2}.
\end{Definition}
\begin{Note}\rm
The Poisson structures \Ref{PB1t} are certainly different for
different $\cV$ and, therefore, are different from \Ref{PB1}, that was
introduced in the principal model.  However, this previous treatment
may be embedded in the following way. The structures \Ref{PB1t} and
\Ref{PB1} are certainly equivalent if we restrict them to the
functionals of $\At$ that are invariant with respect to the choice of
$\cV$, i.e.~invariant with respect to the transformations
\be\la{gt2}
\At\mapsto \th^{-1} \At \th
\ee
with arbitrary $\th\in G$. These were the gauge transformations in the
principal model, generated by \Ref{cona}. Hence, on the set of
observables of the principal model, the different Poisson structures
coincide. Correspondingly, the action of $H^\x$ and $\Ht^{\x}$ from
\Ref{H} and \Ref{Ht} respectively differs only by the unfolding of
such a gauge transformation.

For the coset model it is important to note, that the gauge freedom
\Ref{gt2} is restricted to $H$-valued matrices $\th$, since only that
part of the constraint \Ref{conat} remains first-class here and
generates gauge transformations. This is part of the result of Theorem
\ref{DirA} below.
\end{Note}

\subsubsection{Solution of the constraints}

Given a set of constraints \Ref{conAt} and a Poisson structure
\Ref{PB1t}, the canonical procedure is due to Dirac
\cite{Dira67}. The constraints are separated into first
and second class constraints, of which the latter are explicitly
solved --- which changes the Poisson bracket into the Dirac bracket
---, whereas the former survive in the final theory.

In the case at hand, the essential part of the constraints is of the
second class, such that the Poisson structure has to be modified and
only a small part of the constraints survives as first-class
constraints. We state the final result as
\begin{Theorem}\la{DirA}
The Dirac procedure for treating the constraints \Ref{conAt} in the
Poisson structure \Ref{PB1t} yields the following Dirac bracket for
the connection $\At$:
\ba\la{PBDt}
\left\{\At^a(\g),\At^b(\m)\right\}_{\cV}^{*} &=& 
-\f12f^{abc}\frac{\At^c(\g)-\At^c(\m)}{\g-\m}\\ 
&&{}+\f12f^{a\n(b)c}\frac{\At^c(\g)}{\m-\f1{\g}}
+\f12f^{\n(a)bc}\frac{\At^c(\m)}{\g-\f1{\m}},\nn
\ea
where the notation of indices means a choice of basis with
$t^{\n(a)}\!\equiv\!\n(t^a)$. The bracket for the logarithmic
derivatives of the conformal factor remains unchanged:
\be\la{cfD}
\Big\{ \x, -(\log h)_\x \Big\}_{\cV}^{*} 
= \Big\{ \xb, -(\log h)_\xb  \Big\}_{\cV}^{*} = 1, 
\ee

The structure is compatible with the
(now strong) identity
\be\la{solvcon}
\At(\g)+\frac1{\g^2}\n\left(\At\left(\frac1{\g}\right)\right) 
= \frac1{\g}\At_\i = \frac1{\g}\n(\At_\i),
\ee
such that compared with \Ref{conAt} it remains the first-class
constraint
\be\la{conQ}
\At_\i+\n(\At_\i) = 0.
\ee

\end{Theorem}
\paragraph{Proof:}
The main idea of the proof is the separation of the variables
$\At(\g)$ into weakly commuting halves: 
\ba
\Phi_1(\g) &:=&
\At(\g)+\frac1{\g^2}\n\left(\At\left(\frac1{\g}\right)\right)
-\frac1{\g}\At_\i \non \Phi_2(\g) &:=&
\At(\g)-\frac1{\g^2}\n\left(\At\left(\frac1{\g}\right)\right)
-\frac1{\g}\At_\i \nn
\ea
with:
\be\la{commu}
\left\{\Phi_1^a(\g),\Phi_2^b(\m)\right\}_{\cV}~\approx~ 0
\ee
on the constraint surface \Ref{conAt}, as follows from \Ref{PB1t} by
direct calculation, using the fact, that $\n$ is an automorphism:
$f^{abc}=f^{\n(a)\n(b)\n(c)}$.

The whole constraint surface is spanned by $\Phi_1=0$ and $\At_\i=0$,
whereas $\Phi_2$ covers the remaining degrees of freedom. Since
$\Phi_1$ and $\Phi_2$ contain respectively $\At_\i\mp\n(\At_\i)$, the
relations \Ref{commu} show, that $\At_\i+\n(\At_\i)$ is a first-class
constraint of the theory.

If we further explicitly solve the second-class constraints
$\Phi_1=0$, the commutativity \Ref{commu} implies, that the Poisson
bracket of $\Phi_2$ remains unchanged by the Dirac procedure:
\ben
\left\{\Phi_2^a(\g),\Phi_2^b(\m)\right\}_{\cV}^{*}
=\left\{\Phi_2^a(\g),\Phi_2^b(\m)\right\}_{\cV}
\een
Moreover, the Dirac bracket is by construction compatible with the
vanishing of $\Phi_1$:
\ben
\left\{\Phi_1^a(\g),~.~\right\}_{\cV}^{*}=0
\een
These facts may be used to easily calculate the Dirac bracket of the
original variables $\At(\g)$ without explicitly inverting any matrix
of constraint brackets. With the decomposition
\ben
\At(\g)=\frac12\Phi_1(\g)+\frac12\Phi_2(\g) 
+\frac1{2\g}(\At_\i+\n(\At_\i)) +\frac1{2\g}(\At_\i-\n(\At_\i))
\een 
the result is obtained. The bracket \Ref{cfD} follows from the
calculations performed in Lemma \ref{tind}, which imply the vanishing
Poisson bracket between $(\log h)_\x$ and the constraints.
\qed

\subsubsection{Final formulation and symmetries of the theory}\la{fs}
Let us summarize the final status of the theory and the relation of
the new fundamental variables $\At(\g)$ to the original fields $\cV$
and $g$ respectively. We further discuss, how the local and global
symmetries of the original fields become manifest in this formulation.

The formulation in terms of the new variables $\At(\g)$ is completely
described in Theorem \ref{DirA}, where their modified Poisson structure
is given. The solved constraints \Ref{solvcon} may be considered to be
valid strongly.

The remaining first-class constraint \Ref{conQ} generates the
transformation
\be\la{gtH}
\At\mapsto \chi^{-1} \At \chi
\ee
with $\chi\in H$. According to \Ref{cur2}, the field $\cV$ transforms
as
\be\la{gtHV}
\cV\mapsto \chi \cV
\ee

The relation \Ref{constrV} on the coset constraint surface shows, that
the field $g$ does not feel this transformation. The gauge
transformations generated by \Ref{conQ} are the manifestation of a
really physical gauge freedom in the decomposition of the metric into
some vielbein; they are remnant of the gauge freedom of local Lorentz
transformations in general relativity. This freedom may be fixed to
choose some special gauge for the vielbein field $\cV$.

\begin{Note}\rm
It is important to notice, that the second term in the modified
Hamiltonians $\Ht^{\x}, \Ht^{\xb}$ from \Ref{Ht}, that makes them
differ from $H^{\x}, H^{\xb}$ from \Ref{H} becomes pure gauge
generator after the presented solution of the constraints. This is due
to the fact, that $\At_\i\in{\mathfrak h}$ according to
\Ref{solvcon}. Since ${\mathfrak h}$ and ${\mathfrak k}$ are
orthogonal with respect to the Cartan-Killing form, the action of
$H^{\x}$ and $\Ht^{\x}$ just differs by ${\mathfrak h}$-conjugation
and thus by a gauge transformation of the coset model.
\end{Note}

The field $\At$ now does not contain the complete information about
the original field $\cV$, but only the currents $\cV_\x \cV^{-1},
\cV_\xb \cV^{-1}$, which may be extracted from $\At(\pm 1)$ by means
of \Ref{cur2}. At first sight, one might get the impression, that in
contrast to \Ref{cur}, the relations \Ref{cur2} do not even contain
the full information about these currents. However, if the gauge
freedom \Ref{gtHV} in $\cV$ is fixed, the currents may be uniquely
recovered from \Ref{cur2}. For ${\mathfrak g}=\mathfrak{sl}(N)$ for
example, usually a triangular gauge of $\cV$ is chosen, such that
$\cV_\x \cV^{-1}$ is recovered from its symmetric part $2P_+ = (\cV_\x
\cV^{-1})\!+\!(\cV_\x \cV^{-1})^t$.

The field $\cV$ moreover is determined only up to right multiplication
$\cV\mapsto \cV\th$ from the currents $\cV_\x \cV^{-1}, \cV_\xb
\cV^{-1}$. This is a (global) symmetry of the theory, under which the
field $g$ according to \Ref{constrV} transforms as:
\be\la{Ehlers}
g\mapsto \th^{-1}g\n(\th)
\ee
For axisymmetric stationary $4D$ gravity these are the so-called Ehlers
transformations. They are obviously a symmetry of the original
equations of motion \Ref{ee}.

The new variables $\At(\g)$ are invariant under these global
transformations, which become only manifest in the transition to the
original fields. The related $\Psit$-function transforms due to its
normalization at $\i$ as
\be
\Psit\mapsto\Psit\th\ee
as well as the auxiliary matrix $C_0$, which is related to
$\Psit(\g\!=\!0)$:
\be
C_0\mapsto\n(\th)^{-1}C_0\th.
\ee

Thereby, we have made explicit the global and local symmetries of the
original fields in the new framework.

\subsubsection{First order poles}
Let us evolve the previous result for the case of simple poles of
$\At(\g)$. We again parametrize $\At(\g)$ by its singularities and
residues:
\be\la{Atpar}
\At(\g)=\sum_{j=1}^N \f{\At_j}{\g-\g_j}
\ee
Thus
\be
\At_j= \cV A_j \cV^{-1}
\ee
Their equations of motion read:
\ba
\f{\p \At_j }{ \p \x } =
\f{2}{\x-\xb}\sum_{k\neq j}\f{[\At_k,\;\At_j]}{(1-\g_k)(1-\g_j)}+
[\cV_{\x} \cV^{-1}, \At_j]
\la{1t}\\
\f{\p \At_j }{ \p \xb } =
\f{2}{\xb-\x}\sum_{k\neq j}\f{[\At_k,\;\At_j]}{(1+\g_k)(1+\g_j)}
+[\cV_{\xb} \cV^{-1}, \At_j]\nn
\ea
and are completely generated by the Hamiltonians $\Ht^{\x}$ and
$\Ht^{\xb}$ from \Ref{Ht}.

Theorem \ref{DirA} now implies
\begin{Corollary}
Let $\At$ be parametrized as in \Ref{Atpar}.  After the Dirac
procedure, the following identities hold strongly:
\be
\g_j=\f{1}{\g_{j+n}}
\la{constrg}
\ee
\be
\At_j=\eta(\At_{j+n})\nn
\la{constrAj}\ee
where $N\!=\!2n$. They may be explicitly checked to also commute with
the full Hamiltonian constraints $\Cx, \Cxb$. The remaining
degrees of freedom are therefore covered by the $\g_j$ and $\At_j$ for
$1\!\le\!j\!\le\!n$, which are equipped with the Dirac bracket:
\be
\left\{\At_i^a,\At_j^b\right\}_{\cV}^{*} = 
\frac12\d_{ij}f^{abc}\At_j^c
\ee
The remaining first-class constraint is
\be
\frac12 \left(\At_\i+\n(\At_\i)\right) = 
\sum_{j=1}^{n} \At_j + \n\left(\sum_{j=1}^{n} \At_j \right) = 0
\ee
\end{Corollary}
\qed

This solution of the constraints in the case of first order poles
may alternatively be carried out in terms of the monodromies $M_j$.
As was mentioned above, in the presence of only simple poles, the
variables $A_j$ are generically (see Note \ref{ms}) completely defined
by the monodromies $M_j$.

Assuming that \Ref{constrg} is fulfilled, the coset constraints in the
form \Ref{conPsit} are equivalent to
\be\la{consM}
M_{j+n} - C_0^{-1} \eta(M_j) C_0 = 0.
\ee

There are two important points that this form of the constraints
exhibits. First, it shows the necessity to choose the matrix $C_0$ to
be constant in the isomonodromic sector. Moreover, it uniquely relates
the ordering of the monodromy matrices fixed for calculation of its
Poisson brackets in Theorem \ref{mono} to the ordering defined by
\Ref{constrg}. This results from choosing the corresponding paths
pairwise symmetric under $\g\!\mapsto\!\frac1{\g}$.

The goal is now to calculate the Dirac bracket between
monodromies $M_j$ with respect to \Ref{constrAj}, or, equivalently,
with respect to \Ref{consM}. One way is clearly to repeat the
calculation of Theorem \ref{mono} using the Dirac bracket \Ref{PBDt}
instead of the Poisson bracket \Ref{PB1}.
However, we can alternatively determine the Dirac bracket from simple
symmetry arguments avoiding direct calculation at least for objects
that are invariant under $G$-valued gauge transformations (i.e.~traces
of arbitrary products of $M_j$). 

The involution $\n^{\i}$ introduced by \Ref{eta} acts on $M_j$ 
according to \Ref{conPsit} as follows:
\be
\n^{\i}(M_j)= C_0 \n(M_{j+n}) C_0^{-1}
\la{invM}\ee
Therefore, the set of all $G$-invariant functionals of $M_j$
may be represented as
\be
M_S \oplus M_{AS} 
\ee
where the set $M_S$ contains functionals which are invariant with
respect to $\n^{\i}$ and $M_{AS}$ contains functionals changing the
sign under the action of $\n^{\i}$. Since $\n$ is an automorphism of
the structure \Ref{monoMiMi}, \Ref{monoMiMj}, the definition of
$\n^{\i}$ in \Ref{invM} implies, taking into account Note \ref{etat}:
\be
\{M_S, M_{S}\}\subseteq M_S\qquad
\{M_S, M_{AS}\}\subseteq M_{AS}\qquad
\{M_{AS}, M_{AS}\}\subseteq M_S
\ee
The constraints \Ref{consM} are equivalent to vanishing of all
functionals from $M_{AS}$; therefore the part of $G$-invariant
variables surviving after the Dirac procedure is contained in $M_S$.
The former Poisson bracket on $M_S$ coincides with
the Dirac bracket.

\begin{Note}\rm
The treatment of coset constraints in terms of the monodromies
presented above is invariant with respect to change of $\cV$ since the
monodromies of $\Psit$ are. Therefore, this treatment also works in
the former Poisson structure \Ref{PB1}.
\end{Note}

\subsection{Quantum coset model}
The quantization of the coset model goes along the same line as the
quantization of the principal model described above. We again restrict
to the first order pole sector of the theory, although generalization
to the whole isomonodromic sector should be achievable according to
Note \ref{qhipo}.

Having solved the constraints, the remaining degrees of freedom are
the singularities $\g_j$, the residues $\At_j$ for $j\!=\!1,\dots,n$
and the logarithmic derivatives of the conformal factor $h$. They may
be represented as in \Ref{hq} and \Ref{Aj} again. The quantum
representation space is 
\ben
V^{(n)}:=V_1\otimes\dots\otimes V_n
\een

The Wheeler-De Witt equations \Ref{WDW1} take the form:
\ba\la{WDWcos}
\frac{\p\psi}{\p\x} &=& 
\frac{i\hbar}{\x-\xb}
\left\{\sum_{j,k}\frac{1+\g_j\g_k}
{(1-\g_j)(1-\g_k)}\,\O_{jk} - 
\sum_{j,k}\frac{\g_j+\g_k}
{(1-\g_j)(1-\g_k)}\,\tilde{\O}_{jk}\right\} \psi \\
\frac{\p\psi}{\p\xb} &=& 
\frac{i\hbar}{\xb-\x}
\left\{\sum_{j,k}\frac{1+\g_j\g_k}
{(1+\g_j)(1+\g_k)}\,\O_{jk} + 
\sum_{j,k}\frac{\g_j+\g_k}
{(1+\g_j)(1+\g_k)}\,\tilde{\O}_{jk}\right\}\psi \nn
\ea
with
\ben
\O_{jk} = t^a_j\otimes t^a_k\qquad 
\tilde{\O}_{jk} := t^{\n(a)}_j\otimes t^a_k
\een

Additionally, the physical states have to be annihilated by the
first-class constraint \Ref{conQ}:
\be\la{conQq}
\left(\sum_j t_j^a+\sum_j t_j^{\n(a)}\right) \psi(\x,\xb) = 0
\ee

The result of Theorem \ref{KZWDW} is modified to establish a link to
solutions of what we will refer to as the {\bf Coset-KZ-system}:
\be\la{KZcos}
\frac{\p\varphi_{\scriptscriptstyle CKZ}}{\p\g_j} = 
i\hbar\left\{\sum_{k\not=j}\frac{1+\g_k/\g_j}
{\g_j-\g_k}\,\O_{jk} + 
\sum_{k}\frac{\g_k+1/\g_j}
{\g_j\g_k-1}\,\tilde{\O}_{jk}\right\}\varphi_{\scriptscriptstyle CKZ}
\ee

The relation between solutions of the Wheeler De-Witt equations and
solutions of the Coset-KZ-system is now explicitly given by
\begin{Theorem}\la{KZWDWcos}
If $\varphi_{\scriptscriptstyle CKZ}$ is a solution of \Ref{KZcos}
obeying the constraint \Ref{conQq}, and the $\g_j$ depend on
$(\x,\xb)$ according to \Ref{gamma}, then
\be\la{linkcos}
\psi = 
\prod_{j=1}^{n} 
\left( \g_j^{-1}\f{\p\g_j}{\p w_j}\right)^{i\hbar\O_{jj}} 
\varphi_{\scriptscriptstyle CKZ} 
\ee
solves the constraint (Wheeler-DeWitt) equations \Ref{WDWcos}. 
\end{Theorem}
This may directly be calculated in analogy to \Ref{link}.\qed

The procedure of identifying observables may be outlined just as
in the case of the principal model, where this was described in great
detail. Again the monodromies of the quantum linear system are the
natural candidates for building observables and contain a complete set
for the simple pole sector. In analogy to Theorem \ref{ResT} they
should be identified with the monodromies of a certain
higher-dimensional Coset-KZ-system with an additional insertion
playing the role of the classical $\g$. The actual observables are
generated from combinations of matrix entries of these monodromies
that commute with the constraint \Ref{conQq}. From general reasoning
according to the classical procedure, relevant objects turn out to be
the combinations of $G$-invariant objects, that are also invariant
under the involution $\n_\i$.

\subsection{Application to dimensionally reduced Einstein gravity}
Let us finally sketch, how the previous formalism and results work
for the case of axisymmetric stationary $4D$ gravity. In this case,
the Lagrangian of general relativity is known to reduce to
\Ref{Lagrangian} with the field $g$ taking values in $\G$ as a
symmetric $2\times2$ matrix; its symmetry corresponds to the coset
constraint \Ref{coset}. 

Most of the physically reasonable solutions of the classical theory
--- among them in particular the Kerr solution --- lie in the
isomonodromic sector and are described by first order poles at purely
imaginary singularities in the connection. The quantization of this
sector may be performed within the framework of this paper. According
to \Ref{Aj} and Note \ref{ising} the residues $\At_j$ are represented
as:
\be
\At_j\equiv i \hbar \pmatrix{\frac12h_j & e_j \cr f_j & -\frac12h_j \cr},
\la{Aquant}\ee
where $h_j, e_j$ and $f_j$ are the Chevalley generators of $\ga$.

Due to its non-compactness, $\ga$ admits no finite dimensional unitary
representations, but several series of infinite dimensional
representations. The study of the classical limit singles out the
principal series, as was discussed in \ci{KorNic96}. The
representation space consists of complex functions $f(\z)$ on the real
line with the ordinary $L^2(\R)$ scalar product:
\be
\langle f_1,f_2 \rangle := \int_\R\overline{f_1(\z)}f_2(\z){\rm d}\z
\ee
and the anti-hermitean operators act as
\be
h_j \equiv 2\z_j\p_j + s_j,\qquad e_j \equiv \z_j^2\p_j+s_j\z_j,\qquad
f_j\equiv -\p_j
\ee
The spin $s_j$ takes values $s_j\!=\!1\!+\!iq_j$ with a continuous
parameter $q_j\in\R$.

The surviving first-class constraint \Ref{conQq} now takes a simple form:
\begin{Lemma}\la{red}
A solution $f(\z_1,\dots,\z_n)$ of the constraint \Ref{conQq} is of
the form
\be\la{form}
f(\z_1,\dots,\z_n) = \prod_j(\z_j^2+1)^{-\frac12s_j}
F(\tilde{\z}_1,\dots,\tilde{\z}_n)
\ee
with $\tilde{\z}_j:=\frac{\z_j+i}{\z_j-i}$ and
\be
\left(\sum_j \f{\p}{\p \tilde{\z}_j}\right) F = 0
\ee
\end{Lemma}
This follows by direct calculation.\qed The prefactor in \Ref{form} is
exactly sufficient for convergence of the integral, such that for
finiteness of the norm, it is sufficient to demand boundedness of $F$
which is a function on the product of $(n-1)$ circles $S^1$. In
contrast to the analogous $\ga$ representation of the principal model,
where solutions of finite norm are absent due to several redundant
integration variables, a convergency factor here comes out for
free. This interestingly resembles the fact, that the general reason
for dividing out the maximal compact subgroup in the physical coset
models corresponds to avoiding unboundedness of the energy in the
theory.

It remains to solve the Coset-KZ-system in this
representation. Although the general solution for $\ga$ is not known,
one might be able to obtain explicit results for a small number of
insertions. The Kerr solution for instance, which is of major
interest, requires only two classical insertions $\g_1, \g_2\in
i\R$. In this case, we may exploit Theorem \ref{KZWDWcos} and Lemma
\ref{red} to explicitly reduce the WDW equation to a second order
differential equation in two variables. Let $V_1$ and $V_2$ be two
representations from the principal series of $\ga$ fixed by $s_1$ and
$s_2$ and parametrize the quantum state $\psi(\x,\xb)\in V_1\otimes
V_2$ according to:
\be
\psi(\x,\xb,\z_1,\z_2) ~=~
(\z_1^2+1)^{-{\frac12s_1}} (\z_2^2+1)^{-{\frac12s_2}}
\left(\frac{\g_1}{\g_1^2-1}\right)^{\D_1}
\left(\frac{\g_2}{\g_2^2-1}\right)^{\D_2}\;
F(\g,\z)
\ee
with
\ben
\D_1\equiv\frac{i}2\hbar s_1(s_1-2)\,,\quad
\D_2\equiv\frac{i}2\hbar s_2(s_2-2)\,,
\een
\ben
\g\equiv\frac{\g_1+1}{\g_1-1}\frac{\g_2-1}{\g_2+1}\,\in S^1,\qquad
\z\equiv\frac{\z_1+i}{\z_1-i}\frac{\z_2-i}{\z_2+i}\,\in S^1.
\een
After some calculation the WDW equation then becomes:
\be\la{wd}
\p_\g F(\g,\z) ~=~ i\hbar D_{s_1,s_2}(\g)\;\; F(\g,\z)
\ee
with
\ba 
D_{s_1,s_2}(\g) &=& \left\{
\frac1{\g-1}
\left[2\z(\z\!-\!1)^2\p_\z^2+
\left(2(\z\!-\!1)^2+(s_1+s_2)(\z^2\!-\!1)\right)\p_\z
+\frac{\z^2\!+\!1}{2\z}s_1s_2\right]\right. \non
&&{}\hspace{-0.5em}- \frac1{\g+1}
\left[2\z(\z\!+\!1)^2\p_\z^2+
\left(2(\z\!+\!1)^2+(s_1+s_2)(\z^2\!-\!1)\right)\p_\z
+\frac{\z^2\!+\!1}{2\z}s_1s_2\right] \non
&& {}\hspace{-0.5em} \left.
+ \frac4{\g}\,(\z^2\p^2_\z+\z\p_\z)\; \right\}
\ea
This form e.g.~suggests expansion into a Laurent series in $\z$ on $S^1$
leading to recurrent differential equations in $\g$ for the
coefficients. Further study of this equation should be a subject of
future work. 
\begin{Note}\rm
Equation \Ref{wd} reduces to a Painlev\'e equation when the
principal series representation of $\ga$ is formally replaced by the
fundamental representation of $\mathfrak{g}\!=\!\mathfrak{su}(2)$.  In
the study of four-point correlation-functions in Liouville theory a
similar generalization of the hypergeometric differential equation
appeared \ci{FatZam86}.  
\end{Note}

\section{Outlook}

We have completed the classical two-time Hamiltonian formulation of
the coset model for the isomonodromic sector and sketched a continuous
extension in Appendix \ref{beyi}.  For the quantum theory it remains
the problem of consistent quantization of the total phase space
including a proper understanding of the structures \Ref{beyM}.
The most important physical problem in the investigated model is the
description of states corresponding to quantum black holes. One may
certainly hope to extract first insights from a closer study of the
exact isomonodromic quantum states of the coset model identified in
the last chapter, in particular from the study of equation \Ref{wd}.
\medskip

An open problem is the link of the employed two-time Hamiltonian
formalism with the conventional one. To rigorously relate the
different Poisson structures, the canonical approach should be
compared to our model after a Wick rotation into the Lorentzian
case. This corresponds to a dimensional reduction of spatial
dimensions only, such that the model would describe colliding plane or
cylindrical waves rather than stationary black holes.  It is further
reasonable to suspect that proper comparison of the different Poisson
structures can only be made on the set of observables, see also Note
\ref{comPB}. Recent progress in the canonical approach has been
stated in \ci{KorSam96}, where in particular the canonical algebraic
structures of the observables have been revealed. However, so far the
canonical and the isomonodromic approaches appear to favor different
characteristic observables, which still remain to be related.

As another possibility to compare our treatment with canonical
approaches, the relation to further restricted and already studied
models should be investigated. Of major interest in this context would
be for instance the relation to the Einstein-Rosen solutions,
investigated and quantized in \ci{Kuch71, AshPie96}, where imposing of
additional hypersurface orthogonality of the Killing vector fields
reduces the phase space to ``one polarization'', yet maintaining an
infinite number of degrees of freedom. 

An additional interesting field of future research descends from the
link to broadly studied two-dimensional dilaton gravity (see
e.g.~\ci{CGHS92,GeKuLo95,BarKun96,Fili96}), further allowing to extract
information about the black hole thermodynamics. Further relevance of
the investigated model appeared in certain sectors of string theory
\ci{Galt95,Maha95}.

\paragraph{Acknowledgments:} 
It is a pleasure to thank H.\,Nicolai, V.\,Schomerus and J.\,Teschner
for enlightening discussions. D.K. acknowledges support of Deutsche
Forschungsgemeinschaft under contract No.~Ni~290/5-1.  H.S. thanks
Studienstiftung des Deutschen Volkes for financial support.
\bigskip
\bigskip

\begin{appendix}

\section{Extension beyond the isomonodromic sector}\la{beyi}

The treatment of the isomonodromic sector presented in this paper
allows rather natural extension to the full phase space. This general
scheme reminds a continuous version of the simple pole sector treated
in subsection \ref{fop}, which in turn may be understood as a discrete
embedding into the former. We will again first describe the scheme for
the principal model and then discuss the modifications required for
the coset model, see also \ci{NiKoSa96}.

\subsection{Principal model}

We start from a simply-connected domain $\O$ in the $\x, \xb$-plane,
symmetric with respect to conjugation $\x\mapsto\xb$, where the
classical solution $g(\x,\xb)$ is assumed to be non-singular. This
regularity is reflected by corresponding properties of the related
$\Psi$-function in the $w$-plane. It is holomorphic and invertible in
a (ring-like) domain $D$ of the Riemann surface ${\cal L}$ of the
function $\sqrt{(w-\x)(w-\xb)}$ bounded by contours $l$ and
$l^\sigma$, where $\sigma$ is the involution $\g\mapsto 1/\g$
interchanging the $w$-sheets of ${\cal L}$.

To simplify the following formulas we further assume the spectral
parameter current $A(\g)$ to be holomorphic on the whole second sheet
of ${\cal L}$, such that it may be represented inside of $l$ (we
denote this simply-connected domain by $D_0$) by a Cauchy integral
over $l$:
\be
A(\mu)=\oint_{l}\f{\A(w,\x,\xb)d w}{\g(w)-\mu},
\la{Cauchy}\ee
which is the continuous analog of the simple pole ansatz \Ref{simplep}
in the isomonodromic sector; $\A(w), w\in l$ is a density
corresponding to the residues $A_j$ from \Ref{simplep}.

From \Ref{Cauchy}, $\A(w)$ is not uniquely defined by the values of
$A(\g), \g\in D_0$, in particular, it may not coincide with the
boundary values of $A(\g)$ on $l$. To fix $\A(w)$, we postulate the
following deformation equations which are a continuous version of the
discrete deformation equations \Ref{1}:
\ben
\f{\p \A(w)}{\p\x}=
\f{2}{\x-\xb}\oint_{l}
\f{[\A(v),\;\A(w)]}{(1-\g(v))(1-\g(w))} dv
\een
\be
\f{\p \A(w)}{\p\xb}=
\f{2}{\xb-\x}\oint_{l}
\f{[\A(v),\;\A(w)]}{(1+\g(v))(1+\g(w))} dv\;\;\;\;\;
w\in l
\la{contdef}\ee

It is easy to check that \Ref{contdef} together with \Ref{Cauchy} 
imply the deformation equations \Ref{A1} for $A(\g)$.

The Poisson structure on $\A(w)$ is also a direct continuous analog of   
\Ref{PBAsp}:
\be
\{\A^a(w), \A^b(v)\}= -f^{abc} \A^c(w) \delta (w-v)\qquad
w,v\in l,
\la{PBAc}\ee
where $\delta (w)$ is a one-dimensional $\delta$-function living
on the contour $l$ (and should, strictly speaking, be defined as
$\frac{{\rm d}s}{{\rm d}w} \delta(s)$ with an arbitrary affine
parameter $s$ along $l$).  This structure in turn induces the proper
holomorphic bracket \Ref{PB1} for $A(\g)$:
\ba
\{A^a(\g(w)), A^b (\g(v))\} &=&
-f^{abc}\oint_l\f{\A^c(w')d w'}{(\g(w')-\g(w))
(\g(w')-\g(v))} \non
&=& -f^{abc}\f{A^c\g((w))-A^c(\g(v))}{\g(w)-\g(v)} \nn
\ea

The nice feature of $\A(w)$ in contrast to $A(\g)$ is that $\A(w)$ (as
its discrete analog $A_j$) is explicitly $(\x,\xb)$ independent,
i.e.~the whole dependence of $\A(w)$ on $\x$ and $\xb$ is generated by
the Hamiltonians \Ref{H} (note that the points $\g=\pm 1$ lie inside
of $D_0$):
\be
H^\x=\f{1}{\x-\xb}\tr\left[\oint_{l}\f{\A(w) dw}{1-\g(w)}\right]^2
,\qquad
H^\xb=\f{1}{\xb-\x}\tr\left[\oint_{l}\f{\A(w) dw}{1+\g(w)}\right]^2
\la{Hcont}\ee

We may now also identify a continuous family of observables,
generalizing the construction of section \ref{algob}.  Define $A(\g)$
inside and outside of $D_0$ by the Cauchy formula \Ref{Cauchy} and
construct the related functions $\Psi_{in}(\g\!\in\!D_0)$ and
$\Psi_{out} (\g\!\not\in\!D_0)$ according to $\Psi_\g\Psi^{-1}=
A(\g)$. Then the continuous monodromy matrix
\be
M(w)\equiv \Psi_{out}(w)\Psi^{-1}_{in}(w)\;\;,\;\;\; w\in l
\ee
is $\x,\xb$-independent, since both $\Psi_{in}$ and $\Psi_{out}$
satisfy the linear system \Ref{ls}. Calculations similar to those in
Appendix \ref{pmono} yield the following Poisson brackets for $M(w)$
\ba
\{M^0(v),M^{\bar 0}(w)\} &=& 
i\pi \,\Big(- M^0(v)\,\O\,M^{\bar 0}(w) + M^{\bar 0}(w)\,\O\,M^0(v)  
\label{monoMuMv}\\
&&{}+\O\,M^0(v) M^{\bar 0}(w) - M^0(v) M^{\bar 0}(w)\,\O\Big)\qquad 
v\leq w,\quad v,w\in l\nn
\ea
where the points of contour $l$ are ordered with respect to a fixed point
$w_0$, playing the role of the eyelash in the discrete case.

The brackets \Ref{monoMuMv}, are again valid up to the first-class
constraint generated by
\be
A_\i = \oint_{l} \A(w) dw
\ee
and therefore satisfy Jacobi identities only being restricted to the
gauge-invariant objects.

Again there appear two fundamental ways of quantization. In terms of
$\A$, \Ref{PBAc} would be replaced by a possibly centrally extended
affine algebra. Alternatively, the Poisson algebra of observables
\Ref{monoMuMv} may be quantized directly after regularization
analogously to \Ref{monorMM}:
\ba
\{M^0(v),M^{\bar 0}(w)\} &=& 
- M^0(v)\,r_+\,M^{\bar 0}(w) + M^{\bar 0}(w)\,r_-\,M^0(v) \non  
&&{}+r_-\,M^0(v) M^{\bar 0}(w) - M^0(v) M^{\bar 0}(w)\,r_+ \qquad
v\leq w,\quad v,w\in l\nn 
\ea
leading to:
\be\la{beyM}
R_-M^0(w)R_-^{-1}M^{\bar 0}(v) = 
M^{\bar 0}(v)R_+M^0(w)R_+^{-1},\qquad v\leq w
\ee
\medskip

Embedding of the isomonodromic sector into the presented extension looks
especially simple if all the singularities $\g_1,\dots, \g_N$ 
are assumed to belong to the contour $l$. The density $\A(w)$ is then
parametrized as 
\be
\A (w) =-\sum_{j=N}^n A_j \delta(w-w_j)
\la{Acvd}\ee
where the residues $A_j$ are the same as in \Ref{Atpar}. The Poisson
structure \Ref{PBAc} is the directly inherited from \Ref{PBAsp}
and \Ref{Acvd}:
\ba
\{\A^a(w),\;\A^b(v)\} &=& 
\sum_{j=1}^{N}f^{abc} A_j\delta(w-w_j)\delta(v-w_j)\non
&=& -f^{abc}\A^c(v)\delta(v-w)\nn
\ea
The monodromy $M(w)$ here is a step function on $l$ with jumps at $w=w_j$.
Fixing the eyelash between $\g_N$ and $\g_1$ it is
\ben
M(w) = M_1\dots M_j\qquad w\in ]\g_j,\g_{j+1}[
\een
\begin{Note}\rm
Isomonodromic solutions with higher order poles are embedded into the
general scheme by inserting higher order derivatives of $d$-functions
into \Ref{Acvd}. The definition \Ref{Cauchy} already shows, that the
proper object in this case is the connection $A^w=\frac{\p\g}{\p w}A$,
in accordance with the results from subsection \ref{hp}.
\end{Note}

\begin{Note}\rm 
The representation \Ref{Cauchy} gains a well known meaning, when the
model is truncated to a real scalar field $g$, where $\A(w)$ becomes
independent of $\x,\xb$ and the equation of motion \Ref{ee} reduces to
the Euler-Darboux equation
\be
\p_{\x}\p_{\xb}\phi -\f{\p_{\x}\phi-\p_{\xb}\phi}
{2(\x-\xb)} = 0
\la{ED}\ee
for $\phi=\log g$. Solutions of this equation may be represented as
\ci{CouHil31}
\be
\phi=\oint_l \f{ f(w) dw}{\sqrt{(w-\x)(w-\xb)}}
\la{SED}\ee
with $2\pi i f(w)\equiv\phi(\x\!=\!\xb\!=\!w)$ defined on the axis
$\x=\x$ and continued analytically. After differentiating in $\x$ and
integrating by parts in $w$, this representation takes the form
\ben
\p_\x \phi = \f{2}{\x-\xb}\oint_l\f{f(w) dw}{\sqrt{(w-\x)(w-\xb)}}
\een
and thus equals \Ref{cur} with $A(\pm 1)$ defined by \Ref{Cauchy}
after identification of $f'(w)$ and $\A(w)$.
\end{Note}

\subsection{Coset model}

In analogy to the discrete case, the coset model is most conveniently
described in terms of modified variables 
\ben
\cAt=\eta(\cV) \A \eta(\cV^{-1})
\een
Due to the symmetry \Ref{conAt} between the values of $\At(\g)$ on
different sheets of ${\cal L}$, we can no longer assume $\At(\g)$ to
be holomorphic in $D_0$, but have to replace the $l$ by $l\cup l^\s$
enclosing $D$ in the formulas of the last section.
The coset constraints in terms of $\cAt (w)$ take the form
\be
\cAt(w)=\eta\big(\cAt(w^\sigma)\big)\qquad w\in l
\la{contcoset}\ee
and allow rather simple solution via a Dirac procedure, such that the
phase space is reduced to the values of $\cAt (w)$ on $l$ only,
equipped with the Dirac bracket
\be
\{\cAt^a(w), \cAt^b(v)\}^*_{\cV}=-\f{1}{2} f^{abc} \cAt^c(w) 
\delta (w-v)\;\;\;\;\;\; v,w\in l
\la{CCDB}\ee
Via the Cauchy representation \Ref{Cauchy} on the contour $l\cup
l^\s$, this bracket further gives the previously derived Dirac bracket
\Ref{PBDt} on $\At(\g)$.  It remains the ${\mathfrak h}$-valued first
class constraint
\ben
\oint_l \Big(\cAt (w) +\eta (\cAt (w))\Big)dw =0,
\een
generalizing \Ref{conQ}. The Hamiltonians finally also take the form
\Ref{Hcont} with $l$ being replaced by $l\cup l^\s$. 
In terms of the observables $M(w)$, restriction to the coset leads to
\ben
M(w^\sigma) = C_0^{-1} \eta\Big(M(w)\Big) C_0\;\;\;\;\;\; w\in l
\een
with some constant matrix $C_0$ playing the same role as in \Ref{consM}.

\section{Poisson structure of monodromy matrices}\label{pmono}

This appendix is devoted to the proof of Theorem \ref{mono}, which was
obtained in collaboration with H.\,Nicolai.\footnote{After
completion we learned about related results in \ci{AleMal96,Hitc97}.}
For simplicity of the presentation, we give the calculation for the
case, where the Casimir element $\O$ differs from the permutation
operator $\Pi$ by some scalar multiple of the identity only, which is
the case for ${\mathfrak g} = \mathfrak{sl}(N,\R)$ for example. The
procedure may easily be extended (concerning the notation mainly) to
the general case.

In this case, the Poisson-structure of the connection is given by
\ben
\{A(\g){\,\stackrel \otimes ,\, }A(\m)\} = 
\frac1{\g-\m}\,[\Pi, A(\g)\otimes I + I \otimes A(\m)]
\een

and the statement to be proven reads:
\ba
\{M_i {\,\stackrel \otimes ,\, } M_i\} &=& 
i\pi\, [\, \Pi, M_iM_i\otimes I \,] \label{monoMiMiA}\\
\{M_i {\,\stackrel \otimes ,\, } M_j\} &=& 
i\pi \Pi \,\Big( M_jM_i\otimes I + I\otimes M_iM_j - 
M_i\otimes M_j - M_j\otimes M_i \Big) \label{monoMiMjA} \\
&& {\it for }\enspace i<j \nn
\ea

We first calculate the Poisson structure of matrix entries of the
function $\Psi$ at different points $s_1$ and $s_2$. These points are
defined on the Riemann surface given by $\Psi$ by paths, connecting
them to a common base-point $s_0$, at which $\Psi$ is taken to be
normalized according to (\ref{norm}). The limit $s_0\!\rightarrow\!
\infty$ will be treated later on.

For the calculation, we make use of the standard formula:
\ba
\{\Psi(s_1) {\,\stackrel \otimes ,\, } \Psi(s_2)\} &=&
\Big(\Psi(s_1) \otimes \Psi(s_2)\Big)
          \int_{s_0}^{s_1}\int_{s_0}^{s_2} 
                                     {\rm d}\m_1 {\rm d}\m_2
                                       \times\non
&&\hspace{3em}
\Big(\Psi^{-1}(\m_1) \otimes \Psi^{-1}(\m_2)\Big)
\,\Big\{A(\m_1) {\,\stackrel \otimes ,\, }A(\m_2) \Big\}\,
\Big(\Psi(\m_1)\otimes \Psi(\m_2)\Big) \nn , 
\ea
where the integrand may be rewritten as
\ben
\frac{\Pi}{\m_2-\m_1}\enspace
\Big( \partial_{\m_1} + \partial_{\m_2} \Big) 
\Big( \Psi^{-1}(\m_2)\Psi(\m_1)\otimes\Psi^{-1}(\m_1)\Psi(\m_2)\Big) 
\een

\begin{figure}[htbp]
  \begin{center}
    \leavevmode
     \input{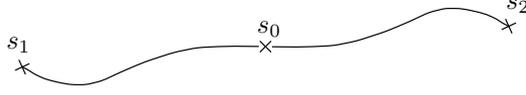}
  \end{center}
  \caption{Choice of paths}
  \label{line}
\end{figure}

This expression is completely regular, even for $\m_1=\m_2$. However,
if the appearance of the derivation operators is exploited by partial
integration, the integrals will split up into parts that exhibit
singularities in coinciding points $\m_1=\m_2$. Thus, we restrict to
distinguished endpoints $s_1$ and $s_2$, choosing the defining paths
$[s_0\!\rightarrow\! s_1]$ and $[s_0\!\rightarrow\! s_2]$
nonintersecting in the punctured plane from the very
beginning. Singularities remain in the common endpoints of the paths
at~$s_0$. As a regularization, one of these coinciding endpoints is
shifted by a small (complex) amount~$\e$ that is put to zero
afterwards. Then, partial integration can be carried out properly,
leaving only boundary terms, that lead to surviving simple line
integrals, whereas the remaining double integrals cancel exactly.  The
arising singularities in $\e=0$ regularize each other such that the
sum is independent of the way, $\e$ tends to zero. In a comprehensive
form, the result may be stated as
\begin{Theorem} \label{PoissonPsi}

Let $s_1$ and $s_2$ be different points on the punctured plane,
defined as points on the covering by nonintersecting paths
$[s_0\!\rightarrow\! s_1]$ and $[s_0\!\rightarrow\! s_2]$ with common
basepoint $s_0$ at which $\Psi$ is normalized. Then, the Poisson
bracket between matrix entries of $\Psi(s_1)$ and $\Psi(s_2)$ is given
by
\ba
\{\Psi(s_1) {\,\stackrel \otimes ,\, } \Psi(s_2)\} 
&=&
\Big(\Psi(s_1) \otimes \Psi(s_2)\Big) \enspace\times 
\label{poissonpsi}\\
&&\left\{\hspace*{0.5em} 
\int_{s_0}^{s_2}{\rm d}\m\,
\frac{\Pi}{\m-s_1}\enspace
\Big( \Psi^{-1}(\m)\Psi(s_1)\otimes\Psi^{-1}(s_1)\Psi(\m)\Big)\right.
\non
&&{} -
\int_{s_0}^{s_1}{\rm d}\m\,
\frac{\Pi}{\m-s_2}\enspace
\Big( \Psi^{-1}(s_2)\Psi(\m)\otimes\Psi^{-1}(\m)\Psi(s_2)\Big)\non
&&{}+
\int_{s_0}^{s_2}{\rm d}\m\,\frac1{\m-s_0}\enspace
\Big[\Pi\, ,\,\Psi(\m)\otimes\Psi^{-1}(\m)\Big] \non
&&{}\left.+
\enspace\lim_{\e\rightarrow 0} 
\left( \int^{s_0-\e}_{s_2} + \int_{s_0+\e}^{s_1} \right){\rm d}\m\,
\frac{\Pi}{\m-s_0}\enspace 
\Big(\Psi(\m)\otimes\Psi^{-1}(\m)\Big) \enspace\right\}.\nn 
\ea 
This expression is regular and independent of the limit procedure.
\qed
\end{Theorem}
\begin{Note} \label{PoissonPsin} \rm
The result of the regularization is the complete fixing of the
relative directions of the paths $[s_0\!\rightarrow\! s_1]$ and
$[s_0\!\rightarrow\! s_2]$ approaching the basepoint $s_0$, that is
determined by the form in which $\e$ arises in the last term in
(\ref{poissonpsi}). In other words, the path
$[s_1\!\rightarrow\!s_0\!\rightarrow\! s_2]$ must pass through the
basepoint $s_0$ straightforwardly, as is indicated in
figure~\ref{line}.
\end{Note}

The result of theorem \ref{PoissonPsi} may be further simplified in
the limit $s_0\!\rightarrow\!\infty$, where the third term of
(\ref{poissonpsi}) vanishes:

\begin{Lemma}  For a fixed point $s$ on the punctured plane and 
$\Psi(\g)$ holomorphic at $\g=\infty$, it is
\be
\lim_{s_0\rightarrow\infty} \left(\, \int_{s_0}^s {\rm d}\m\,
\frac1{\m-s_0}\enspace
\Big[\Pi, \Psi(\m)\otimes\Psi^{-1}(\m)\Big] \,\right) \enspace=
\enspace 0 .
\label{Tvanish}
\ee
\end{Lemma}
The proof is obtained by estimating the integrand as holomorphic
function of $\g$ and $s_0$.
\qed

To proceed in calculating the Poisson bracket between monodromy
matrices, we choose points $s_1, s_2, s_3$ and $s_4$, pairwise
coinciding on the punctured plane as $s_1 \sim s_2$ and $s_3 \sim
s_4$, but distinguished on the covering and defining the monodromy
matrices $M_i$ and $M_j$:

\be
\Psi(s_2) = \Psi(s_1)M_i \qquad \Psi(s_4) = \Psi(s_3)M_j 
\ee  

Then, (\ref{poissonpsi}) leads to:

\ba
\{M_i {\,\stackrel \otimes ,\, } M_j \} &=& 
\hspace*{0.5em}(M_i \otimes M_j)\left[ 
\int_{s_4\!\rightarrow\!s_0\!\rightarrow\!s_2} {\rm d}\m\,
\frac{\Pi}{\m-s_0}\enspace 
\Big(\Psi(\m)\otimes\Psi^{-1}(\m)\Big)\right]\label{monodromies}\\
&&{}+ \hspace*{4em}\left[
\int_{s_3\!\rightarrow\!s_0\!\rightarrow\!s_1}  {\rm d}\m\,
\frac{\Pi}{\m-s_0}\enspace 
\Big(\Psi(\m)\otimes\Psi^{-1}(\m)\Big) 
\right](M_i \otimes M_j) ,\non
&&{}-
(I\otimes M_j) \left[
\int_{s_4\!\rightarrow\!s_0\!\rightarrow\!s_1} {\rm d}\m\,
\frac{\Pi}{\m-s_0}\enspace 
\Big(\Psi(\m)\otimes\Psi^{-1}(\m)\Big) 
\right](M_i\otimes I) \non
&&{}-
(M_i\otimes I)\left[
\int_{s_3\!\rightarrow\!s_0\!\rightarrow\!s_2} {\rm d}\m\,
\frac{\Pi}{\m-s_0}\enspace 
\Big(\Psi(\m)\otimes\Psi^{-1}(\m)\Big) 
\right](I\otimes M_j) \nn
\ea
which is understood in the limit $\e\!\rightarrow\!0$ and
$s_0\!\rightarrow\!\infty$ and for paths
$[s_j\!\rightarrow\!s_0\!\rightarrow\!s_i]\,$, $i=1,2; j=3,4$, chosen
fixed and in accordance with the conditions of Theorem
\ref{PoissonPsi} and Note \ref{PoissonPsin}. 

\begin{figure}[htbp]
  \begin{center}
    \leavevmode
     \input{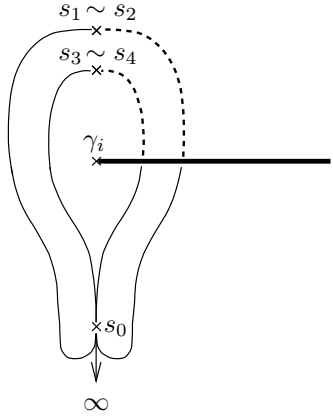}
  \end{center}
  \caption{Choice of paths for $\{M_i {\,\stackrel \otimes ,\, } 
   M_i \}$}
  \label{mono1}
\end{figure}

\paragraph{Proof of (\ref{monoMiMiA}):} Consider first the case
$i=j$. Then a proper choice of paths is illustrated in
figure~\ref{mono1}.

The expression (\ref{monodromies}) allows to put $s_1=s_3$ and
$s_2=s_4$ and to split the integration paths into paths encircling
$s_0$ and $\g_i$, respectively:
\ba
\{M_i {\,\stackrel \otimes ,\, } M_i \} &=& 
(M_i\otimes M_i) X - X (M_i\otimes M_i)  
- (M_i\otimes I)X(I\otimes M_i) \non
&&{}+ (I\otimes M_i)X(M_i\otimes I) + (I\otimes M_i)Y(M_i\otimes I)
    - (M_i\otimes I)Y(I\otimes M_i) , \nn
\ea
with
\ba
X &=&\frac12 \oint_{s_0} {\rm d}\m\,
\frac{\Pi}{\m-s_0}\enspace 
\Big(\Psi(\m)\otimes\Psi^{-1}(\m)\Big) \non
Y &=& \int_{s_1}^{s_2} {\rm d}\m\,
\frac{\Pi}{\m-s_0}\enspace 
\Big(\Psi(\m)\otimes\Psi^{-1}(\m)\Big) \nn
\ea
The path of the integral $Y$ neither passes through $s_0$ nor
intersects the path [$s_0\!\rightarrow\!\infty$]; such that this
integral vanishes in the limit $s_0\!\rightarrow\!\infty$. This choice
of path uniquely determines the orientation of the remaining paths in
$X$, which encircle $s_0$. The corresponding integrals can be easily
evaluated due to Cauchy's theorem and single-valuedness of the
integrands. This proves formula (\ref{monoMiMiA}).  \qed

\begin{figure}[htbp]
  \begin{center}
    \leavevmode
     \input{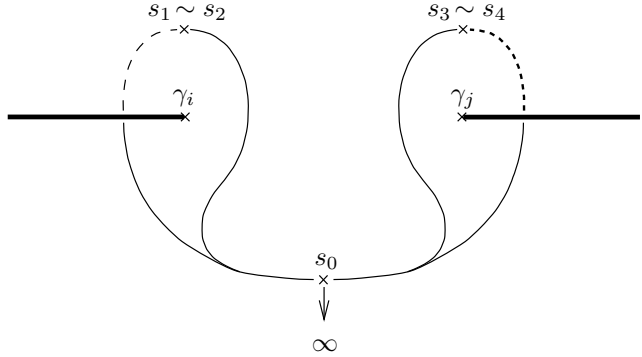}
  \end{center}
  \caption{Paths for $\{M_i {\,\stackrel \otimes ,\, } M_j \}$}
  \label{mono2}
\end{figure}

\paragraph{Proof of (\ref{monoMiMjA}):} This case is treated in
complete analogy. A suitable form of the paths is shown in
figure~\ref{mono2}, which in particular illustrates the asymmetric
position of the paths defining respectively $M_i$ and $M_j$, with
respect to the marked path [$s_0\!\rightarrow\!\infty$].

Similar reasoning as above yields:
\ba
\{M_i {\,\stackrel \otimes ,\, } M_j \} &=&
-(M_i\otimes M_j) X - X (M_i\otimes M_j) \non
&&{}
+ (M_i\otimes I)X(I\otimes M_j) 
+ (I\otimes M_j)X(M_i\otimes I),
\ea
where again several integrals have already vanished in the limit
$s_0\!\rightarrow\!\infty$. Evaluating the remaining terms proves
formula (\ref{monoMiMjA}).
\qed

\end{appendix}

\end{document}